\documentclass[twocolumn]{aastex61}
\usepackage{amssymb}
\usepackage{natbib}
\usepackage{hyperref}
\usepackage{color}
\usepackage{pdfcomment}

\newcommand\changed[1]{{#1}}

\submitjournal{AJ}
\received{2017 November 23}
\accepted{2018 January 3}

\shorttitle{Asteroid Family Associations of Active Asteroids}
\shortauthors{Hsieh et al.}

\begin{document}

\title{Asteroid Family Associations of Active Asteroids}

\correspondingauthor{Henry H.\ Hsieh}
\email{hhsieh@psi.edu}

\author[0000-0001-7225-9271]{Henry H.\ Hsieh}
\affil{Planetary Science Institute, 1700 East Fort Lowell Rd., Suite 106, Tucson, AZ 85719, USA}
\affil{Institute of Astronomy and Astrophysics, Academia Sinica, P.O.\ Box 23-141, Taipei 10617, Taiwan}

\author{Bojan Novakovi\'c}
\affiliation{Department of Astronomy, Faculty of Mathematics, University of Belgrade, Studentski trg 16, 11000 Belgrade, Serbia}

\author{Yoonyoung Kim}
\affiliation{Department of Physics and Astronomy, Seoul National University, Gwanak, Seoul 151-742, Korea}

\author{Ramon Brasser}
\affiliation{Earth-Life Science Institute, Tokyo Institute of Technology, Meguro, Tokyo 152-8550, Japan}


\begin{abstract} 
We report on the results of a systematic search for associated asteroid families for all active asteroids known to date.  We find that 10 out of 12 main-belt comets (MBCs) and 5 out of 7 disrupted asteroids are linked with known or candidate families, rates that have $\sim$0.1\% and $\sim$6\% probabilities, respectively, of occurring by chance, given an overall family association rate of 37\% for asteroids in the inner solar system.  We find previously unidentified family associations between 238P/Read and the candidate Gorchakov family, 311P/PANSTARRS and the candidate Behrens family, 324P/La Sagra and the Alauda family, 354P/LINEAR and the Baptistina family, P/2013 R3-B (Catalina-PANSTARRS) and the Mandragora family, P/2015 X6 (PANSTARRS) and the Aeolia family, P/2016 G1 (PANSTARRS) and the Adeona family, and P/2016 J1-A/B (PANSTARRS) and the Theobalda family.  All MBCs with family associations belong to families that contain asteroids with primitive taxonomic classifications and low average \changed{reported} albedos ($\overline{p_V}$$\,\lesssim\,$0.10), while disrupted asteroids with family associations belong to families that contain asteroids that span wider ranges of taxonomic types and average \changed{reported} albedos (0.06$\,<\,$$\overline{p_V}$$\,<\,$0.25).  These findings are consistent with MBC activity being closely correlated to composition (i.e., whether an object is likely to contain ice), while disrupted asteroid activity is not as sensitive to composition.  Given our results, we describe a sequence of processes by which the formation of young asteroid families could lead to the production of present-day MBCs.
\end{abstract}






\section{INTRODUCTION}\label{section:introduction}
\subsection{Background}\label{section:background}

Active asteroids are small solar system bodies that exhibit comet-like mass loss yet occupy dynamically asteroidal orbits, typically defined as having Tisserand parameters of $T_J$$\,>\,$3.00 and semimajor axes less than that of Jupiter \citep[cf.][]{jewitt2015_actvasts_ast4}.  They include main-belt comets (MBCs), whose activity is thought to be driven by the sublimation of volatile ices \citep[cf.][]{hsieh2006_mbcs}, and disrupted asteroids, whose mass loss is due to disruptive processes such as impacts or rotational destabilization \citep[cf.][]{hsieh2012_scheila}.

Active asteroids have attracted considerable attention since their discovery for various reasons.  MBCs may be useful for probing the ice content of the main asteroid belt \citep[e.g.,][]{hsieh2014_mbcsiausproc}, given that dust modeling, confirmation of recurrent activity, or both show that their activity is likely to be driven by the sublimation of volatile ices \citep[cf.][]{hsieh2012_scheila}, while dynamical analyses indicate that many appear to have formed in situ where we see them today \citep[e.g.,][]{haghighipour2009_mbcorigins,hsieh2012_288p,hsieh2012_324p,hsieh2013_p2012t1,hsieh2016_tisserand} \citep[or if they are originally from the outer solar system, at least must have been implanted at their current locations at very early times; e.g.,][]{levison2009_tnocontamination,vokrouhlicky2016_tnocapturemainbelt}.  No spectroscopic confirmation of sublimation products have been obtained for any MBC studied to date \citep[cf.][]{jewitt2015_actvasts_ast4}, but this lack of direct detections of gas only indicates that gas production rates were below the detection limits of the observations in question at the time, not that gas was definitively absent \citep[cf.][]{hsieh2016_mbcsiausproc}.  MBCs and the evidence they provide of likely present-day ice in the asteroid belt are especially interesting for solar system formation models and astrobiology given dynamical studies that suggest that a large portion of the Earth's current water inventory could have been supplied by the accretion of icy objects either from the outer asteroid belt or from more distant regions of the solar system that were scattered onto Earth-impacting orbits \citep[e.g.,][]{morbidelli2000_earthwater,raymond2004_earthwater,obrien2006_earthwater,raymond2017_waterorigin}.

Meanwhile, disrupted asteroids provide opportunities to study disruption processes for which significant theoretical and laboratory work has been done \citep[e.g.,][]{ballouz2015_impactsimulations,durda2015_disruptionfragments,housen2018_impactsporousasteroids}, but for which real-world and real-time observations are relatively lacking.  Disruption events represent opportunities to probe the structure and composition of asteroid interiors that are difficult to study otherwise \citep[e.g.,][]{bodewits2014_scheila,hirabayashi2014_p2013r3}.

Approximately 20 active asteroids have been discovered to date, although the exact number reported by different sources within the community can vary due to slight differences in dynamical definitions (e.g., the use of different $T_J$ values as the ``asteroidal'' cut-off, such as $T_J$$\,=\,$3.05 or $T_J$$\,=\,$3.08, or the inclusion of objects not confined to the main asteroid belt).  In this paper, we only consider active asteroids whose orbits do not cross those of Mars and Jupiter, have semimajor axes between 4J:1A and 2J:1A mean-motion resonances (MMRs) at 2.065~AU and 3.278~AU (the canonical boundaries of the main asteroid belt), respectively, and have Tisserand parameters with respect to Jupiter of $T_J$$\,>\,$3.

\setlength{\tabcolsep}{4.0pt}
\setlength{\extrarowheight}{0em}
\begin{table*}[htb!]
\caption{Physical and Dynamical Properties of Known Active Asteroids}
\smallskip
\footnotesize
\begin{tabular}{lcrccccrl}
\hline\hline
\multicolumn{1}{c}{Object}
 & \multicolumn{1}{c}{Type$^a$}
 & \multicolumn{1}{c}{$r_N$$^b$}
 & \multicolumn{1}{c}{$T_J$$^c$}
 & \multicolumn{1}{c}{$a_p$$^d$}
 & \multicolumn{1}{c}{$e_p$$^e$}
 & \multicolumn{1}{c}{$\sin(i_p)$$^f$}
 & \multicolumn{1}{c}{$t_{ly}$$^g$}
 & \multicolumn{1}{c}{Ref.$^h$}
 \\
\hline
\multicolumn{4}{l}{\it \underline{Sublimation-driven activity}} \\
~~~(1) Ceres                             & S   & 467.6     & 3.310 & 2.767085 & 0.114993 & 0.167721 &  350.9 & [1,2] \\
~~~133P/Elst-Pizarro (P/1996 N2)         & S/R & 1.9       & 3.184 & 3.163972 & 0.153470 & 0.024165 &  934.6 & [3,4] \\ 
~~~176P/LINEAR ((118401) 1999 RE$_{70}$) & S?  & 2.0       & 3.166 & 3.217864 & 0.145566 & 0.024465 &   92.5 & [4,5] \\
~~~238P/Read (P/2005 U1)                 & S   & 0.4       & 3.153 & 3.179053 & 0.209260 & 0.017349 &   16.7 & [6,7] \\
~~~259P/Garradd (P/2008 R1)              & S   & 0.3       & 3.217 & 2.729305 & 0.280882 & 0.288213 &   33.8 & [8,9] \\
~~~288P/(300163) 2006 VW$_{139}$         & S   & 1.3       & 3.204 & 3.053612 & 0.160159 & 0.037982 & 1265.8 & [10,11] \\
~~~313P/Gibbs (P/2014 S4)                & S   & 0.5       & 3.132 & 3.152211 & 0.205637 & 0.178835 &   12.0 & [12,13] \\
~~~324P/La Sagra (P/2010 R2)             & S   & 0.6       & 3.100 & 3.099853 & 0.114883 & 0.382057 & 1612.9 & [14,15] \\
~~~358P/PANSTARRS (P/2012 T1)            & S?  & $<$1.3    & 3.135 & 3.160515 & 0.196038 & 0.175636 &    8.5 & [16] \\
~~~P/2013 R3-A (Catalina-PANSTARRS)      & S/R & $\sim$0.2 & 3.184 & 3.030727 & 0.259023 & 0.033973 &    4.2 & [17] \\
~~~P/2013 R3-B (Catalina-PANSTARRS)      & S/R & $\sim$0.2 & 3.184 & 3.029233 & 0.236175 & 0.032950 &    3.6 & [17] \\
~~~P/2015 X6 (PANSTARRS)                 & S/R & $<$1.4    & 3.318 & 2.754716 & 0.163811 & 0.059354 &   91.8 & [18] \\
~~~P/2016 J1-A (PANSTARRS)               & S/R & $<$0.9    & 3.113 & 3.165357 & 0.259628 & 0.249058 &   54.0 & [19,20] \\
~~~P/2016 J1-B (PANSTARRS)               & S/R & $<$0.4    & 3.116 & 3.160171 & 0.259843 & 0.247799 &    8.2 & [19,20] \\
\hline
\multicolumn{4}{l}{\it \underline{Disruption-driven activity}} \\
~~~(493) Griseldis                  & I?  & 20.8      & 3.140 & 3.120841 & 0.144563 & 0.267158 &  529.1 & [21,22] \\
~~~(596) Scheila                    & I   & 79.9      & 3.208 & 2.929386 & 0.197608 & 0.226490 &   13.4 & [23,24] \\
~~~(62412) 2000 SY$_{178}$          & I/R & 5.2       & 3.197 & 3.147701 & 0.111265 & 0.096409 &  121.6 & [25,26] \\
~~~311P/PANSTARRS (P/2013 P5)       & R?  & $<$0.2    & 3.661 & 2.189019 & 0.141820 & 0.094563 &   31.6 & [27,28] \\
~~~331P/Gibbs (P/2012 F5)           & I/R & 0.9       & 3.229 & 3.003859 & 0.022816 & 0.179959 & 6666.7 & [29,30] \\
~~~354P/LINEAR (P/2010 A2)          & I/R & 0.06      & 3.583 & 2.290197 & 0.151754 & 0.097421 &  116.8 & [31,32] \\ 
~~~P/2016 G1 (PANSTARRS)            & I   & $<$0.05   & 3.367 & 2.583930 & 0.169074 & 0.205145 & 1818.2 & [33] \\
\hline
\multicolumn{4}{l}{\it \underline{Unknown activity mechanism}} \\
~~~233P/La Sagra                    & ?   & ---       & 3.081 & 2.985806 & 0.479060 & 0.164666 &    0.1 & [34] \\
~~~348P/PANSTARRS                   & ?   & ---       & 3.062 & 3.146828 & 0.311352 & 0.312174 &    3.0 & [35] \\
\hline
\hline
\end{tabular} \\
$^a$ Type of active asteroid in terms of likely activity driver --- S: sublimation; I: impact; R: rotation; ?: unknown/uncertain. \\
$^b$ Effective nucleus radius, in km. \\
$^c$ Tisserand parameter based on current osculating orbital elements (as of UT 2017 July 1). \\
$^d$ Proper semimajor axis, in AU. \\
$^e$ Proper eccentricity. \\
$^f$ Sine of proper inclination. \\
$^g$ Lyapunov time, in kyr. \\
$^h$ References for object-specific activity mechanism determinations and nucleus size measurements: 
[1] \citet{carry2008_ceres}; [2] \citet{kuppers2014_ceres};
[3] \citet{hsieh2004_133p}; [4] \citet{hsieh2009_albedos};
[5] \citet{hsieh2011_176p}
[6] \citet{hsieh2009_238p}; [7] \citet{hsieh2011_238p};
[8] \citet{maclennan2012_259p}; [9] \citet{hsieh2017_259p};
[10] \citet{hsieh2012_288p}; [11] \citet{agarwal2016_288p};
[12] \citet{jewitt2015_313p1}; [13] \citet{hsieh2015_313p};
[14] \citet{hsieh2014_324p}; [15] \citet{hsieh2015_324p};
[16] \citet{hsieh2013_p2012t1};
[17] \citet{jewitt2014_p2013r3};
[18] \citet{moreno2016_p2015x6};
[19] \citet{moreno2017_p2016j1}; [20] \citet{hui2017_p2016j1};
[21] \citet{masiero2014_neowisealbedos}; [22] \citet{tholen2015_griseldis};
[23] \citet{ishiguro2011_scheila2}; [24] \citet{masiero2012_neowisealbedos};
[25] \citet{masiero2011_neowisealbedos}; [26] \citet{sheppard2015_sy178};
[27] \citet{jewitt2013_311p}; [28] \citet{jewitt2015_311p};
[29] \citet{stevenson2012_331p}; [30] \citet{drahus2015_331p};
[31] \citet{jewitt2010_p2010a2}; [32] \citet{agarwal2013_p2010a2};
[33] \citet{moreno2016_p2016g1};
[34] \citet{mainzer2010_233p};
[35] \citet{wainscoat2017_p2017a2}.
\\
\label{table:aaproperties}
\end{table*}

\subsection{Asteroid family associations}\label{section:familyassociations}

Over the years, many active asteroids have been found to be associated with asteroid families, which are groups of asteroids with similar orbital elements that have been inferred to have formed from the catastrophic fragmentation of single parent bodies at some point in the past \citep{hirayama1918_astfam}.  The first known active asteroid, 133P/Elst-Pizarro, was recognized to be a member of the $\sim$2.5~Gyr-old Themis family \citep{nesvorny2003_dustbands} soon after its discovery \citep[cf.][]{boehnhardt1998_133p}.  Since then, two more active asteroids, 176P/LINEAR and 288P/(300163) 2006 VW$_{139}$, have also been associated with the Themis family \citep{hsieh2009_htp,hsieh2012_288p}.  A fourth active asteroid, 238P/Read, is considered to be a possible former Themis family member whose orbit has dynamically evolved to the point at which it is no longer formally dynamically linked to the family \citep{haghighipour2009_mbcorigins}.  All four of these objects are considered to be MBCs based on dust modeling results, confirmation of recurrent activity, or both \citep[e.g.,][]{boehnhardt1998_133p,hsieh2004_133p,hsieh2010_133p,hsieh2011_176p,hsieh2011_238p,hsieh2012_288p,licandro2013_288p,jewitt2014_133p}.

Other active asteroids have also been associated with other asteroid families including 311P/PANSTARRS, 313P/Gibbs, 354P/LINEAR, 358P/PANSTARRS, and (62412) 2000 SY$_{178}$ \citep[][]{hainaut2012_p2010a2,hsieh2013_p2012t1,hsieh2015_313p,jewitt2013_311p,sheppard2015_sy178}.  However, only some of these associations were formally established using standard family-linking techniques.  Others were simply based on the qualitative similarity of each object's osculating orbital elements to those of a nearby family.

In order to clarify the significance of asteroid family membership to active asteroids, we have conducted a search for family associations for all of the known active asteroids to date, and report the results here.  We also describe the properties of these associated families and discuss the implications of our results.

\section{Family Search Methodology}\label{section:methodology}

Members of asteroid families can be identified from their clustering in proper orbital element space (i.e., proper semimajor axis, $a_p$, proper eccentricities, $e_p$, and proper inclination, $i_p$).  Proper orbital elements are quasi-integrals of motion, where the transient oscillations of osculating orbital elements have been largely removed, making them nearly constant over time.  They are therefore well-suited for identifying stable groupings of objects in dynamical parameter space.

We begin our search for asteroid families associated with known active asteroids by computing synthetic proper orbital elements for each object, using the methodology described by \citet{knezevic2000_synthelements} and \citet{knezevic2003_synthelements}.  Synthetic proper elements are about a factor of 3 more accurate than analytically computed proper elements for objects with low to moderate inclinations and eccentricities \citep[cf.][]{knezevic2017_propelements}, and are also significantly more reliable and useful for identifying asteroid families than analytically computed proper elements for objects at higher inclinations \citep[cf.][]{novakovic2011_highifamilies}.  The results of these computations for the known active asteroids, along with computations of Lyapunov times ($t_{ly}$) to characterize their stability (where objects with $t_{ly}$$\,<\,$10~kyr are typically considered dynamically unstable), are listed in Table~\ref{table:aaproperties}.

To identify clustering of family members in proper element space, we employ the Hierarchical Clustering Method \citep[HCM;][]{zappala1990_hcm,zappala1994_hcm}. The HCM identifies groupings of objects such that each cluster member is closer than a certain threshold distance, $\delta$, from at least one other cluster member, a so-called cut-off ``distance'' ($\delta_c$), which typically has units of velocity (i.e., m~s$^{-1}$).  The traditional application of this method \citep[cf.][]{zappala1990_hcm,zappala1994_hcm,novakovic2011_highifamilies} involves the computation of mutual distances among all asteroids within a selected region of orbital element space, and determination of a so-called quasi-random level (QRL), which is used to determine a statistical significance and the optimum $\delta_c$ value for a given family. Because we are interested in searching for families that include specific objects, i.e., the active asteroids, however, we use a slightly different HCM-like approach that starts from a selected central asteroid.  In this case, the volume of the region of interest in proper element space is not defined a priori, but instead grows around the selected central object as the $\delta_c$ value being considered increases.  Given that this method does not include the determination of a QRL, we need to take a different approach for selecting an appropriate $\delta_c$ value for a family associated with a particular central object.

A plot of the number of asteroids associated with a given central body as a function of $\delta_c$ for an asteroid family is typically characterized by an increase in the number of associated asteroids at small $\delta_c$ values (as members of the family are identified by their close proximity in orbital element space), a ``plateau'' (an interval of $\delta_c$ over which family membership remains nearly constant; mainly seen for families that are very cleanly separated from the background asteroid population in orbital element space), and finally, resumed growth in the number of associated asteroids as increasing $\delta_c$ values begin to incorporate a large fraction of the background population. The most appropriate $\delta_c$ value for a family is typically chosen to include the majority of the asteroids associated with the central body within the plateau region, while excluding the asteroids associated with the central body beyond the plateau region, as those objects are assumed to belong to the background population. A typical example of such a plot is shown in Figure~\ref{figure:family_progression_aeolia}.

For active asteroids that have been previously linked to known families, we perform the HCM-based analysis described above using the previously identified nominal central objects of those families, and test whether each active asteroid becomes linked to its respective family at a reasonable $\delta_c$ value \citep[i.e., less than or comparable to the optimum $\delta_c$ values previously found for those families; e.g., from][]{nesvorny2015_astfam_ast4}.  For those active asteroids that have not been previously linked to known families, we perform the same initial analyses using each active asteroid as the starting central object.  In these cases, if we find that an active asteroid becomes linked with the central object of a known family at a $\delta_c$ value less than or comparable to that family's nominal optimum $\delta_c$ value, we then use that family's previously identified central body as the central body in a follow-up HCM-based analysis to verify that the active asteroid becomes linked with its respective family at a reasonable $\delta_c$ value.  If no link between an active asteroid and a known family is found, but a previously unknown family-like cluster of asteroids is tentatively identified, we perform a similar follow-up HCM-based analysis using the largest body of that candidate family as the central object, and again attempt to verify that the active asteroid becomes linked with the candidate family at a reasonable $\delta_c$ value.

We note that not all family growth plots have features that are as cleanly defined as seen in Figure~\ref{figure:family_progression_aeolia}.  For families in high-density regions of the asteroid belt in orbital element space, the plateau in the family growth plot can be poorly defined, and the most appropriate $\delta_c$ for the family can be difficult to identify, if a family can be determined to exist at all.  As such, selection of the best $\delta_c$ value to define a family often necessarily includes some subjective judgment, and in cases of families found in dense regions of the asteroid belt with overlapping populations of objects, analysis of the physical properties of individual asteroids may be employed to further clarify family membership \citep[e.g.,][]{masiero2013_astfams_neowise}.

In this work, we do not perform detailed analyses of each family in question, many of which have already been analyzed in detail in other works, and others which require dedicated individual investigations beyond the scope of this overview paper.  Rather, we investigate whether active asteroids can be linked with families at reasonably low $\delta_c$ values, using previously determined optimum $\delta_c$ values as benchmarks when available \citep[e.g., from][]{nesvorny2015_pdsastfam}.  Here we note that most active asteroids are km-scale in size or smaller, and may therefore be subject to significant Yarkovsky drift \citep[e.g.,][]{bottke2006_yarkovsky}, \changed{as well as to non-gravitational recoil forces due to asymmetric outgassing \citep[cf.][]{hui2017_activeastsnongrav}}.  As such, it is reasonable to expect that some of these bodies might be found in the outer ``halos'' of their respective families, and therefore in some cases, we may consider active asteroids linked at somewhat larger $\delta_c$ values than are typically used to characterize particular families to still be potential members of those families.

We perform HCM-based analyses for all known active asteroids (as of 2017 June 15) as described above using a synthetic proper element catalog for 524\,216 numbered and multi-opposition asteroids retrieved from the {\it AstDyS} website\footnote{\tt http://hamilton.dm.unipi.it/astdys} on 2017 June 15, where we compute the proper elements of the 17 active asteroids and active asteroid fragments under consideration in this work and also add these to the catalog.

\section{RESULTS}\label{section:results}

\subsection{Overview}\label{section:resultsoverview}

We summarize the results of our search for family associations of the active asteroids in the main asteroid belt known to date in Table~\ref{table:family_associations}.  We find that nearly all of the active asteroids that we investigate here have asteroid family associations, a finding whose significance we discuss further in Section~\ref{section:discussion}.  In the remainder of this section, we consider the individual families for which we have found associated active asteroids, divided into those families associated with MBCs and those associated with disrupted asteroids, and discuss the physical and dynamical properties of those families in the context of the likely physical natures of their associated active asteroids.  For the purposes of this work, classifications of active asteroids as MBCs or disrupted asteroids are based on observational confirmation of recurrent activity or dust modeling indicating prolonged dust emission events \citep[cf.][]{hsieh2012_scheila}, given that no direct confirmation of sublimation has been obtained for any of the MBCs studied to date (cf.\ Section~\ref{section:background}).  The regions of the main asteroid belt in which each associated family is located are listed in Table~\ref{table:family_associations}, where asteroids with $a_p$ between the 4J:1A and 3J:1A MMRs (at 2.064~AU and 2.501~AU, respectively) comprise the inner main belt (IMB), asteroids with $a_p$ between the 3J:1A and 5J:2A MMRs (at 2.501~AU and 2.824~AU, respectively) comprise the middle main belt (MMB), and asteroids with $a_p$ between the 5J:2A and 2J:1A MMRs (at 2.824~AU and 3.277~AU, respectively) comprise the outer main belt (OMB).  In cases where new families or clusters are identified, we emphasize that the reliability of these findings remains to be confirmed, requiring detailed individual analyses that are beyond the scope of this work.  \changed{Similarly, especially in the cases of active asteroids linked with their respective families at relatively large $\delta_c$ values or located near major MMRs, follow-up analyses, such as backward dynamical integrations, may be required to more definitively confirm or rule out the family associations reported here.}  Nonetheless, we report these preliminary findings here in order to highlight potential family associations to investigate in more detail in future work.


\setlength{\tabcolsep}{2.5pt}
\setlength{\extrarowheight}{0em}
\begin{table*}[htb!]
\caption{Family Associations of Known Active Asteroids}
\smallskip
\footnotesize
\begin{tabular}{lcrcccclc}
\hline\hline
\multicolumn{1}{c}{Object}
 & \multicolumn{1}{c}{Family}
 & \multicolumn{1}{c}{$n_{fam}$$^a$}
 & \multicolumn{1}{c}{$\delta_{\rm AA}$$^b$}
 & \multicolumn{1}{c}{$\delta_c$$^c$}
 & \multicolumn{1}{c}{Age$^{d}$}
 & \multicolumn{1}{c}{Region$^{e}$}
 & \multicolumn{1}{c}{$\overline{p_{V}}$$^f$}
 & \multicolumn{1}{c}{Sp.\ Type$^g$} \\
\hline
\multicolumn{4}{l}{\it \underline{Sublimation-driven activity}} \\
~~~(1) Ceres                             & \multicolumn{1}{c}{---} & \multicolumn{1}{c}{---} & \multicolumn{1}{c}{---} & \multicolumn{1}{c}{---} & \multicolumn{1}{c}{---} & MMB & 0.090$\pm$0.003$^h$ (1) & C \\ 
~~~133P/Elst-Pizarro (P/1996 N2)         & Themis         & 4782 &  33 &  60 & 2.5$\pm$1.0~Gyr       & OMB & 0.068$\pm$0.017 (2218)  & B/C   \\ 
~~~...                                   & Beagle         &  148 &  19 &  25 & $<$$\,$10~Myr         & OMB & 0.080$\pm$0.014 (30)    & B/C   \\ 
~~~176P/LINEAR ((118401) 1999 RE$_{70}$) & Themis         & 4782 &  34 &  60 & 2.5$\pm$1.0~Gyr       & OMB & 0.068$\pm$0.017 (2218)  & B/C   \\ 
~~~238P/Read (P/2005 U1)                 & Gorchakov$^i$  &   16 &  45 &  75 & ?                     & OMB & 0.053$\pm$0.012 (7)     & C     \\ 
~~~259P/Garradd (P/2008 R1)              & \multicolumn{1}{c}{---} & \multicolumn{1}{c}{---} & \multicolumn{1}{c}{---} & \multicolumn{1}{c}{---} & \multicolumn{1}{c}{---} & MMB & \multicolumn{1}{c}{---} & \multicolumn{1}{c}{---} \\ 
~~~288P/(300163) 2006 VW$_{139}$         & 288P$^j$       &   11 & n/a &  70 & 7.5$\pm$0.3~Myr       & OMB & 0.090$\pm$0.020 (2)     & C      \\
~~~313P/Gibbs (P/2014 S4)                & Lixiaohua      &  756 &  21 &  45 & $\sim$155~Myr         & OMB & 0.044$\pm$0.009 (367)   & C/D/X  \\ 
~~~324P/La Sagra (P/2010 R2)             & Alauda         & 1294 & 108 & 120 & 640$\pm$50~Myr        & OMB & 0.066$\pm$0.015 (687)   & B/C/X  \\ 
~~~358P/PANSTARRS (P/2012 T1)            & Lixiaohua      &  756 &  13 &  45 & $\sim$155~Myr         & OMB & 0.044$\pm$0.009 (367)   & C/D/X  \\ 
~~~P/2013 R3-A (Catalina-PANSTARRS)      & \multicolumn{1}{c}{---} & \multicolumn{1}{c}{---} & \multicolumn{1}{c}{---} & \multicolumn{1}{c}{---} & \multicolumn{1}{c}{---} & OMB & \multicolumn{1}{c}{---} & \multicolumn{1}{c}{---} \\ 
~~~P/2013 R3-B (Catalina-PANSTARRS)      & Mandragora$^k$ &   30 &  59 &  75 & 290$\pm$20 kyr  & OMB & 0.056$\pm$0.019 (9)     & ?    \\ 
~~~P/2015 X6 (PANSTARRS)                 & Aeolia         &  296 &  36 &  50 & $\sim$100~Myr         & MMB & 0.107$\pm$0.022 (43)    & C/Xe   \\ 
~~~P/2016 J1-A (PANSTARRS)               & Theobalda      &  376 &  23 &  60 & 6.9$\pm$2.3~Myr       & OMB & 0.062$\pm$0.016 (107)   & C/F/X  \\ 
~~~P/2016 J1-B (PANSTARRS)               & ...            &  ... &  30 & ... & ...                   & OMB & ...                     & ...    \\ 
\hline
\multicolumn{4}{l}{\it \underline{Disruption-driven activity}} \\
~~~(493) Griseldis            & \multicolumn{1}{c}{---} & \multicolumn{1}{c}{---} & \multicolumn{1}{c}{---} & \multicolumn{1}{c}{---} & \multicolumn{1}{c}{---} & OMB & 0.081$\pm$0.009 (1) & X \\ 
~~~(596) Scheila              & \multicolumn{1}{c}{---} & \multicolumn{1}{c}{---} & \multicolumn{1}{c}{---} & \multicolumn{1}{c}{---} & \multicolumn{1}{c}{---} & OMB & 0.040$\pm$0.001 (1) & T \\ 
~~~(62412) 2000 SY$_{178}$    & Hygiea     & 4854 &  37 &  60 & 3.2$\pm$0.4~Gyr       & OMB & 0.070$\pm$0.018 (1951)  & B/C/D/X \\ 
~~~311P/PANSTARRS (P/2013 P5) & Behrens$^i$    &   20 &  46 &  45 & ?                     & IMB & 0.248$\pm$0.026 (4)     & Q/S/V   \\ 
~~~331P/Gibbs (P/2012 F5)     & 331P$^l$       &    9 & n/a &  10 & 1.5$\pm$0.1~Myr       & OMB & \multicolumn{1}{c}{?} & Q        \\ 
~~~354P/LINEAR (P/2010 A2)    & Baptistina & 2500 &  43 &  48 & $\sim$100--320~Myr    & IMB & 0.179$\pm$0.056 (581)   & S/X     \\ 
~~~P/2016 G1 (PANSTARRS)      & Adeona     & 2236 &  44 &  50 & 620$\pm$190~Myr       & MMB & 0.060$\pm$0.011 (874)   & Ch      \\ 
\hline
\multicolumn{4}{l}{\it \underline{Unknown activity mechanism}} \\
~~~233P/La Sagra     & \multicolumn{1}{c}{---} & \multicolumn{1}{c}{---} & \multicolumn{1}{c}{---} & \multicolumn{1}{c}{---} & \multicolumn{1}{c}{---} & OMB & \multicolumn{1}{c}{---} & \multicolumn{1}{c}{---} \\ 
~~~348P/PANSTARRS    & \multicolumn{1}{c}{---} & \multicolumn{1}{c}{---} & \multicolumn{1}{c}{---} & \multicolumn{1}{c}{---} & \multicolumn{1}{c}{---} & OMB & \multicolumn{1}{c}{---} & \multicolumn{1}{c}{---} \\ 
\hline
\hline
\end{tabular} \\
$^a$ Number of family members, as computed by \citet{nesvorny2015_pdsastfam}, unless otherwise specified. \\
$^b$ Cut-off distance, in m~s$^{-1}$, at which the specified active asteroid becomes linked with the specified family. \\
$^c$ Cut-off distance, in m~s$^{-1}$, for family in HCM analysis, as determined by \citet{nesvorny2015_pdsastfam}, unless otherwise specified. \\
$^d$ Estimated age of family (?: unknown), from references in text. \\
$^e$ Region of the main asteroid belt in which the specified family is found (IMB: Inner Main Belt; MMB: Middle Main Belt; OMB: Outer Main Belt). \\
$^f$ Average \changed{reported} $V$-band geometric albedos of objects for which values are available; from \citet{mainzer2016_neowise} (?: no albedos available for any known family members). \\
$^g$ Spectral types of family members for which taxonomic classifications are available; from \citet{neese2010_taxonomy} and \citet{hasselmann2011_taxonomy} (?: no classifications available for any known family members). \\
$^h$ \changed{Reported} $V$-band geometric albedo for Ceres determined by \citet{li2006_ceres}. \\
$^i$ Candidate family identified and parameters determined by this work. \\
$^j$ Family identified and parameters determined by \citet{novakovic2012_288p}. \\
$^k$ Family identified and parameters determined by \citet{pravec2017_astclusters}. \\
$^l$ Family identified and parameters determined by \citet{novakovic2014_331p}. \\
\label{table:family_associations}
\end{table*}

\subsection{Main-Belt Comet Family Associations}\label{section:mbcfamilies}


\subsubsection{The Aeolia Family}\label{section:aeolia}

\begin{figure}[htb!]
\centerline{\includegraphics[width=2.6in]{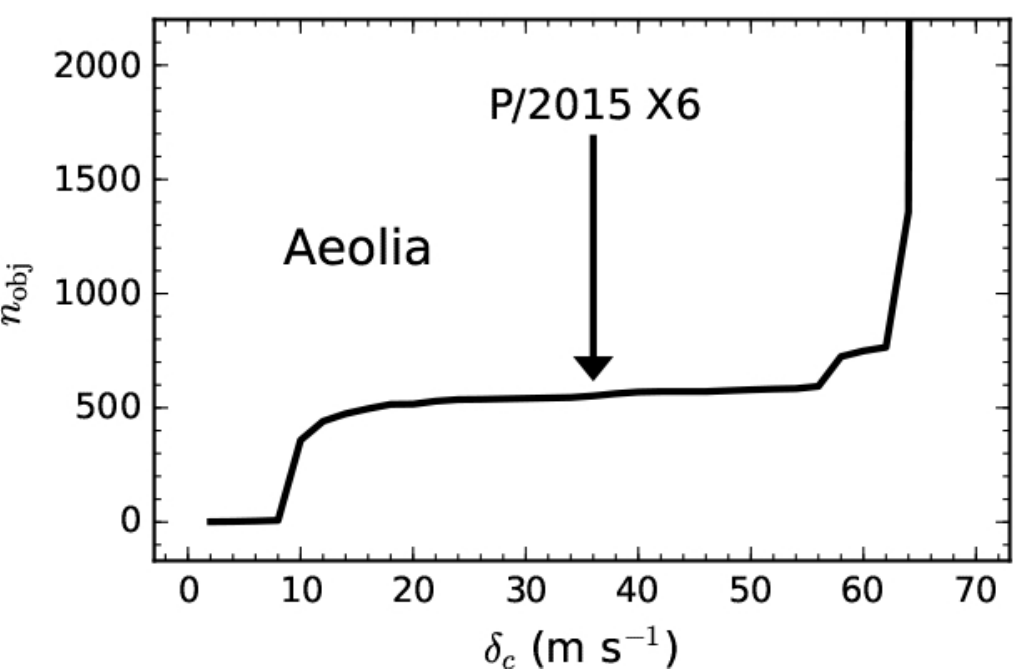}}
\caption{\small Plot of number of asteroids associated with (396) Aeolia as a function of $\delta_c$, where the point at which P/2015 X6 becomes linked with the family ($\delta_c$$\,=\,$36~m~s$^{-1}$) is marked with a vertical arrow.
}
\label{figure:family_progression_aeolia}
\end{figure}

We find that active asteroid P/2015 X6 (PANSTARRS) is linked to the Aeolia family, which is believed to have formed in a cratering event $\sim\,$100~Myr ago \citep{spoto2015_astfamages}.  P/2015 X6 becomes linked with the Aeolia family at $\delta_c$$\,=\,$36~m~s$^{-1}$ (Figure~\ref{figure:family_progression_aeolia}), which is actually outside the optimum cut-off distance ($\delta_{c}$$\,=\,$20~m~s$^{-1}$) determined for the family by \citet{nesvorny2015_astfam_ast4}.  As can be seen in Figure~\ref{figure:family_progression_aeolia} though, P/2015 X6 still lies well within the ``plateau'' region of the family growth plot (cf.\ Section~\ref{section:methodology}) for the Aeolia family, and so we still regard it as a likely family member.

\begin{figure}[htb!]
\centerline{\includegraphics[width=2.1in]{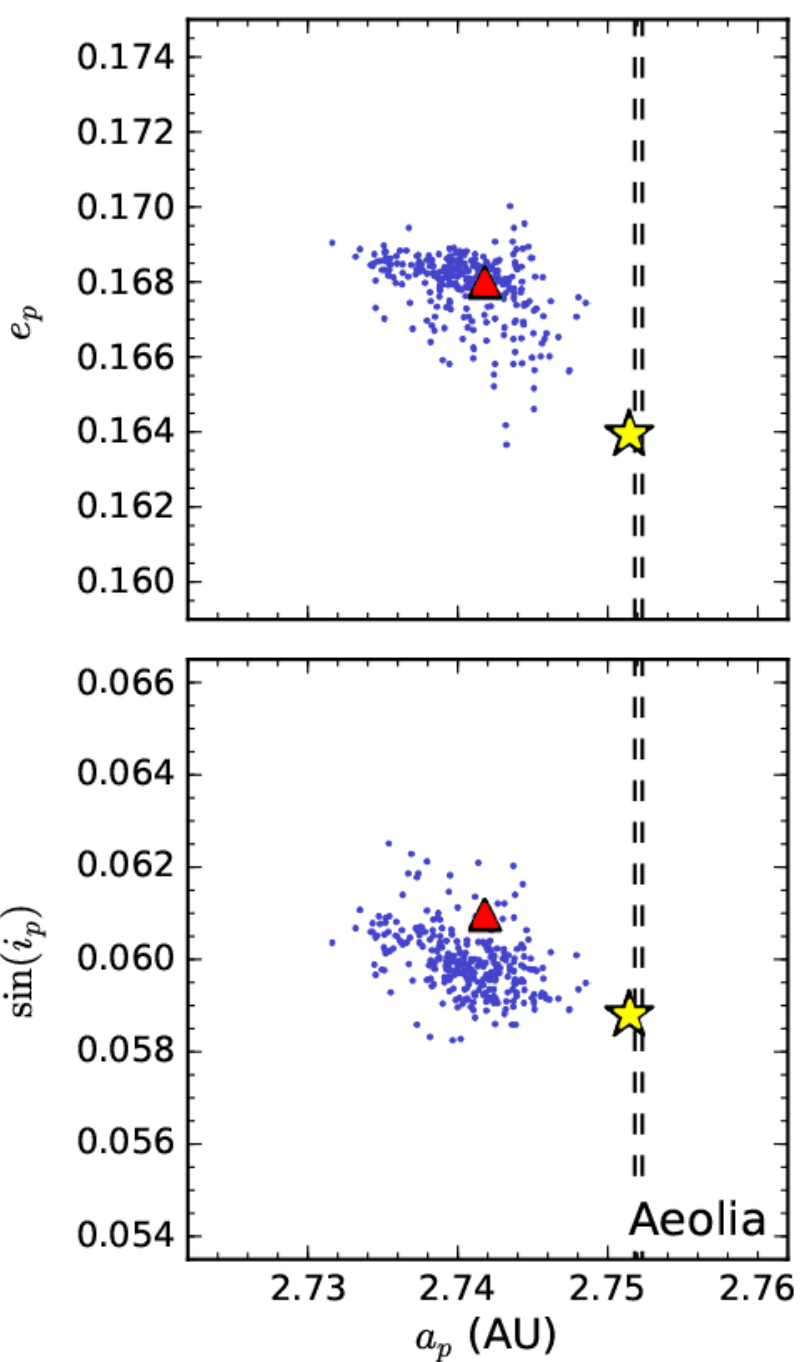}}
\caption{\small Plots of $a_p$ versus $e_p$ (top panel) and $\sin(i_p)$ (bottom panel) for Aeolia family members (small blue dots) identified by \citet{nesvorny2015_pdsastfam}.  The proper elements for (396) Aeolia are marked with red triangles, while the proper elements for P/2015 X6 are marked with yellow stars.
Vertical dashed lines mark the semimajor axis positions of the 3J$-$1S$-$1A (left) and 13J:5A (right) MMRs
at 2.7518~AU and 2.7523~AU, respectively.
}
\label{figure:aei_aeolia}
\end{figure}

The family lies just inside the 13J:5A and 3J$-$1S$-$1A MMRs (Figure~\ref{figure:aei_aeolia}).  P/2015 X6 lies close to those two resonances, indicating that it may be unstable over long timescales.  This potential instability is reflected by its small $t_{ly}$ value (Table~\ref{table:aaproperties}).  Given that P/2015 X6 is also relatively distant in proper element space from the core of the family \citep[and in fact is actually outside the $\delta_c$ cut-off established by][]{nesvorny2015_pdsastfam}, its membership in the Aeolia family may be considered somewhat uncertain.

The largest member of the family, (396) Aeolia, has been spectroscopically classified as a Xe-type asteroid \citep{neese2010_taxonomy}, and \changed{has been reported to have} a geometric albedo of $p_V$$\,=\,$0.126\changed{$\pm$0.019} and effective radius of $r_e$$\,=\,$19.6\changed{$\pm$0.2}~km \citep{mainzer2016_neowise}.  All other family members that have been taxonomically classified have been classified as C-type asteroids.  The average \changed{reported} albedo of Aeolia family members is $\overline{p_{V}}$$\,=\,$0.107$\pm$0.022 (cf.\ Table~\ref{table:family_associations}), although albedos of individual family members \changed{have been reported to} range widely from ${p_V}$$\,\sim\,$0.05 to ${p_V}$$\,\sim\,$0.15, suggesting that the family could have a mix of primitive and non-primitive members \changed{(or alternatively, that individual reported albedo values have large uncertainties)}.

\changed{One important caveat that applies here, as well as to discussions of the physical properties of other families that follow below, is that albedos reported by \citet{mainzer2016_neowise}  (and many others) are generally calculated using absolute $V$-band magnitudes ($H_V$) computed using photometric data compiled by the Minor Planet Center from a wide range of observers and surveys.  However, \citet{pravec2012_wiseabsmagnitudes} found that while catalogued absolute magnitudes for larger asteroids ($H_V$$\,\lesssim\,$10) were generally consistent with results from an independent targeted observing campaign to verify $H_V$ values for several hundred main-belt and near-Earth asteroids, catalogued $H_V$ values for smaller asteroids ($H_V$$\,\gg\,$10) exhibited systematically negative offsets up to $\Delta H_V$$\,\sim\,$$-$0.5 relative to independently measured values.  In many cases, the eventual resulting offsets between catalogued albedo values and recalculated albedo values using revised $H_V$ values were within the originally reported uncertainties of the catalogued albedo values, but nonetheless, we note that albedo values discussed here, particularly for the smaller asteroids that dominate the families we discuss in this paper, should be regarded with some caution.}

A dust modeling analysis of the activity of P/2015 X6 indicates that the object underwent sustained dust loss over a period of at least two months, suggesting that the observed activity was sublimation-driven \citep{moreno2016_p2015x6}, making the object a likely MBC.


\subsubsection{The Alauda Family}\label{section:alauda}

\begin{figure}[htb!]
\centerline{\includegraphics[width=2.6in]{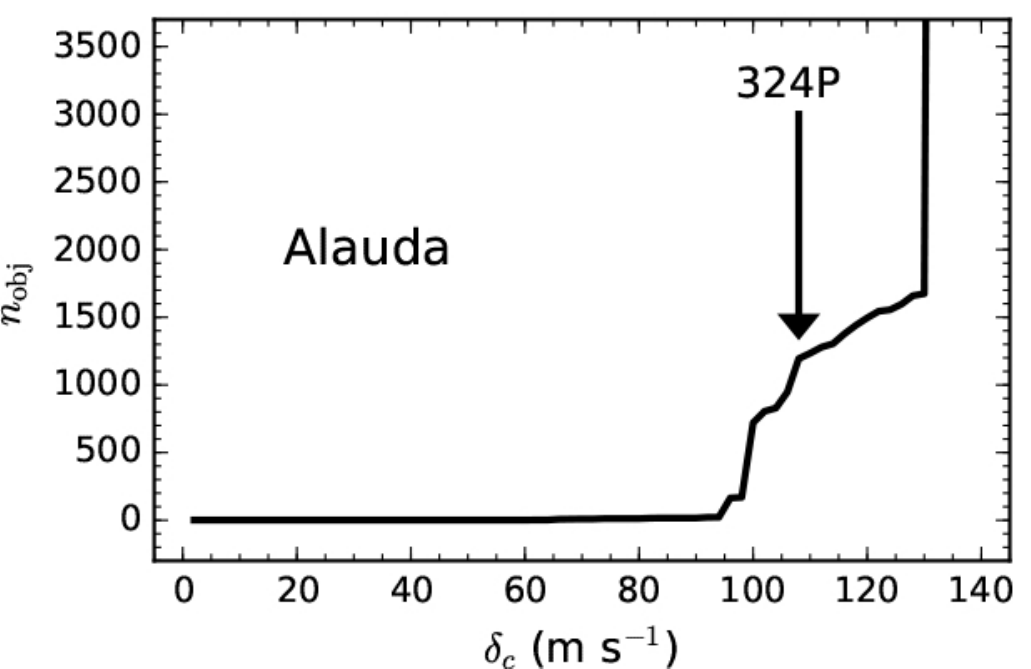}}
\caption{\small Plot of number of asteroids associated with (702) Alauda as a function of $\delta_c$, where the point at which 324P becomes linked with the family ($\delta_c$$\,=\,$108~m~s$^{-1}$) is marked with a vertical arrow.}
\label{figure:family_progression_alauda}
\end{figure}

We find that active asteroid 324P/La Sagra (formerly designated P/2010 R2) is linked to the Alauda family, which has been determined to be 640$\pm$50~Myr old \citep{carruba2016_oldestfamilies}.  324P becomes linked with the Alauda family at $\delta_c$$\,=\,$108~m~s$^{-1}$ (Figure~\ref{figure:family_progression_alauda}), just within the optimum cut-off distance ($\delta_{c}$$\,=\,$120~m~s$^{-1}$) determined for the family by \citet{nesvorny2015_astfam_ast4}.

\begin{figure}[htb!]
\centerline{\includegraphics[width=2.1in]{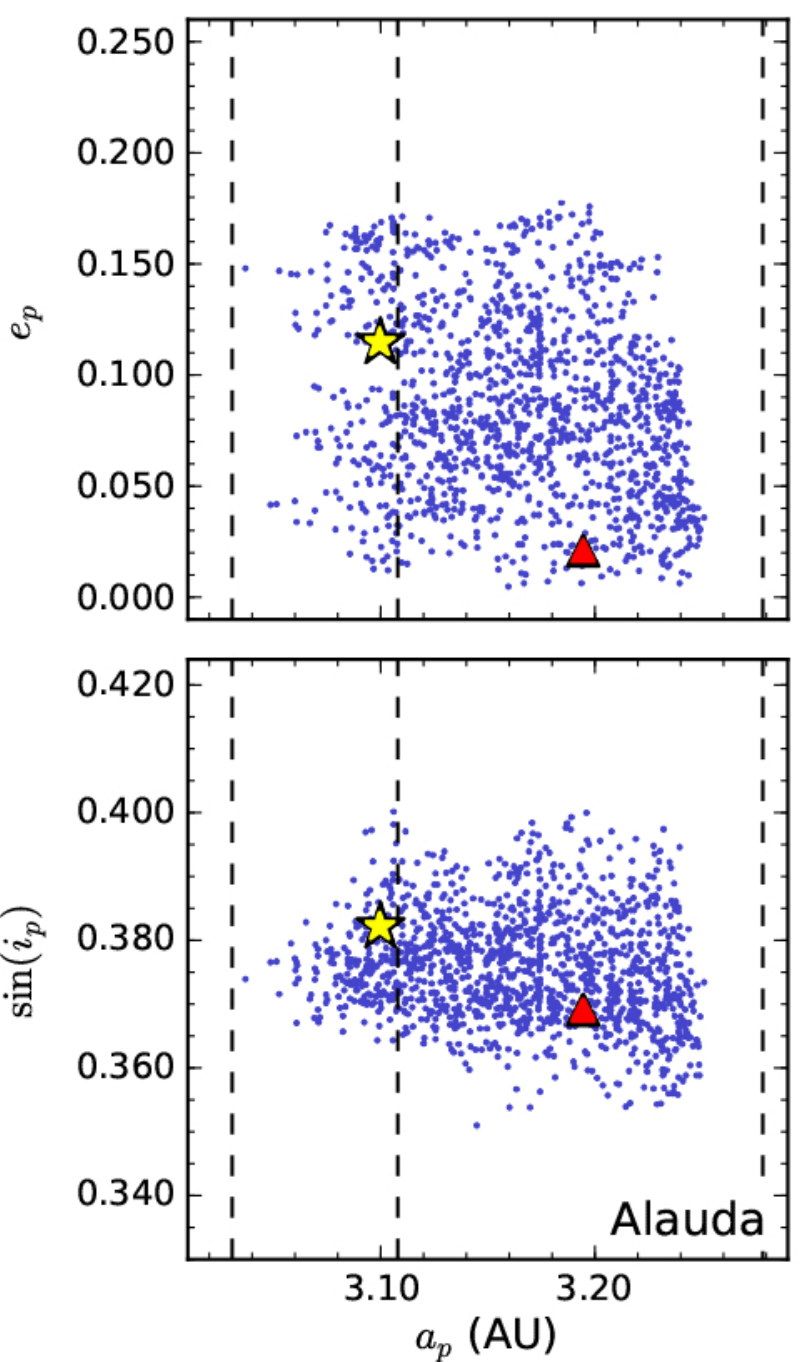}}
\caption{\small Plots of $a_p$ versus $e_p$ (top panel) and $\sin(i_p)$ (bottom panel) for Alauda family members (small blue dots) identified by \citet{nesvorny2015_pdsastfam}.  The proper elements for (702) Alauda are marked with red triangles, while the proper elements for 324P are marked with yellow stars.
Vertical dashed lines mark the semimajor axis positions, from left to right, of the 9J:4A, 13J:6A, and 2J:1A MMRs
at 3.0307~AU, 3.1080~AU, and 3.2783~AU, respectively.
}
\label{figure:aei_alauda}
\end{figure}

The family is found between the 9J:4A and 2J:1A MMRs, and is crossed by the 13J:6A MMR as well as other two- and three-body MMRs (Figure~\ref{figure:aei_alauda}). 
It is bounded above in proper inclination space by the Euphrosyne family and below by the Luthera family, and is also adjacent to the Danae and Erminia families in proper semimajor axis space, separated by the 9J:4A MMR.  Some exchange of objects may be possible between the Alauda family and surrounding families via the $\nu_6$ secular resonance (which connects it to the Danae region), and various three-body MMRs (which connect it to the Euphrosyne and Luthera families) \citep{machuca2012_euphrosyne}.  Several sub-families and clumps within this region have been identified \citep{machuca2012_euphrosyne}, but 324P is not linked to any of them at $\delta_c$ values smaller than the $\delta_c$ value at which it is linked to the main Alauda family.

The largest member of the family, (702) Alauda, has been spectroscopically classified as a B-type asteroid \citep{neese2010_taxonomy}, possesses a small satellite, and \changed{has been reported to have} $p_V$$\,=\,$0.061\changed{$\pm$0.011}, $r_e$$\,=\,$95.5\changed{$\pm$1.0}~km, and a bulk density of $\rho$$\,=\,$1570$\pm$500~kg~m$^{-3}$ \citep{bus2004_alaudaspectrum,rojo2011_alauda,mainzer2016_neowise}.  Other family members have been taxonomically classified as B-, C-, and X-type asteroids, and \changed{have been reported to have} a low average albedo of $\overline{p_V}$$\,=\,$0.066$\pm$0.015 (cf.\ Table~\ref{table:family_associations}), indicating that they are likely to have primitive compositions.

Photometric and morphological analysis of the activity of 324P/La Sagra in 2010 suggested that it was likely to be sublimation-driven \citep{hsieh2012_324p}, a conclusion that was strengthened by the detection of recurrent activity in 2015 \citep{hsieh2015_324p}, making the object a likely MBC.  The object's nucleus has been measured to have $r_e$$\,=\,$0.55$\pm$0.05~km \citep[assuming a $R$-band albedo of $p_R$$\,=\,$0.05;][]{hsieh2014_324p}.


\subsubsection{The Gorchakov Family}\label{section:gorchakov}

\begin{figure}[htb!]
\centerline{\includegraphics[width=2.6in]{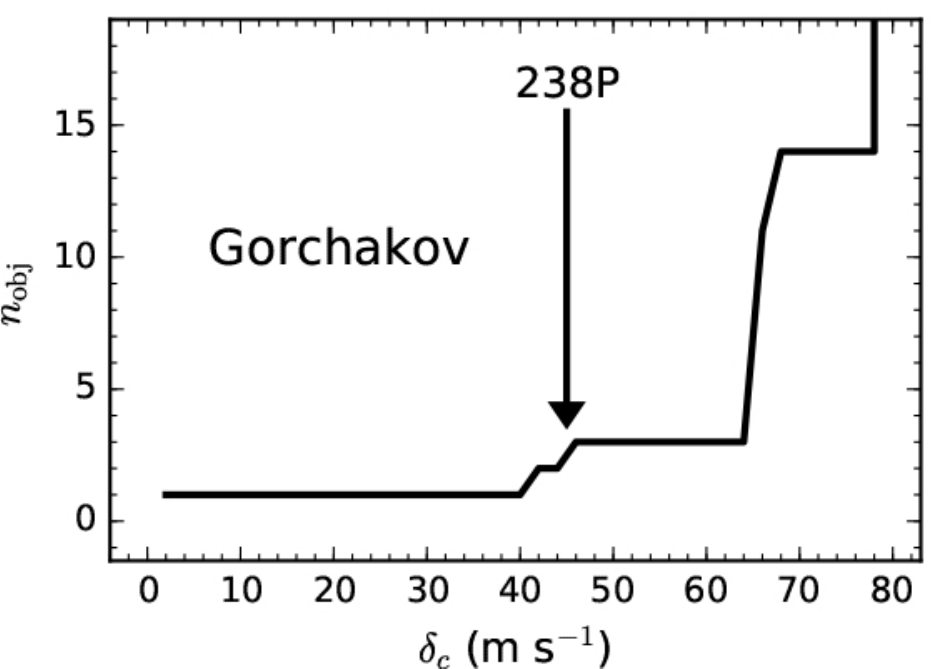}}
\caption{\small Plot of number of asteroids associated with (5014) Gorchakov as a function of $\delta_c$, where the point at which 238P becomes linked with the family ($\delta_c$$\,=\,$45~m~s$^{-1}$) is marked with a vertical arrow.
}
\label{figure:family_progression_gorchakov}
\end{figure}

We find that active asteroid 238P/Read (formerly designated P/2005 U1) is linked to a candidate asteroid family that we designate here as the Gorchakov family.  238P becomes linked with the candidate Gorchakov family at $\delta_c$$\,=\,$45~m~s$^{-1}$ (Figure~\ref{figure:family_progression_gorchakov}).
The largest member of the family, (5014) Gorchakov, \changed{has been reported to have} $p_V$$\,=\,$0.057\changed{$\pm$0.008} and $r_e$$\,=\,$9.7\changed{$\pm$0.1}~km \citep{mainzer2016_neowise}, where its low albedo suggests that it may have a primitive composition.
Other family members have been classified as C-type asteroids, and \changed{have been reported to have} a low average \changed{reported} albedo of $\overline{p_{V}}$$\,=\,$0.053$\pm$0.012 (Table~\ref{table:family_associations}), indicating that they are likely to have primitive compositions.

\begin{figure}[htb!]
\centerline{\includegraphics[width=2.1in]{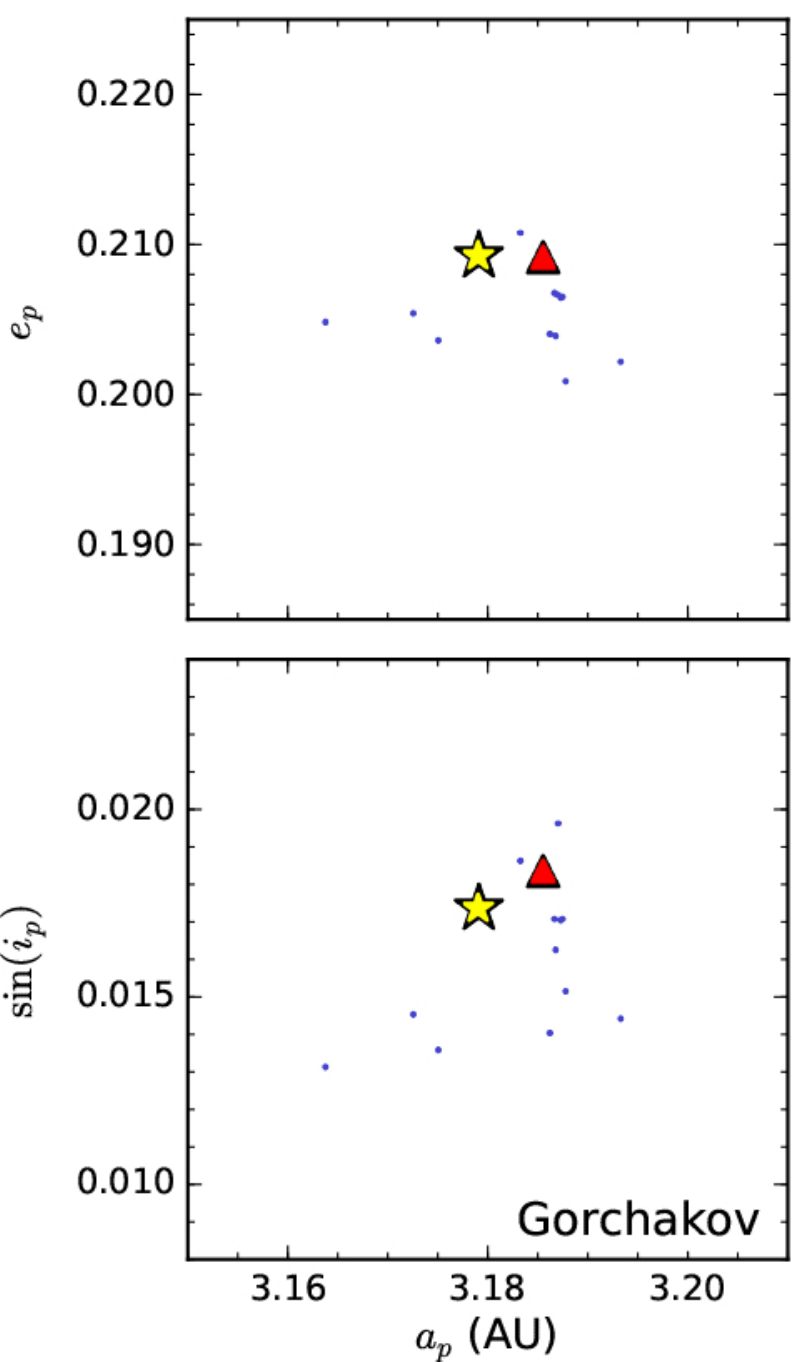}}
\caption{\small Plots of $a_p$ versus $e_p$ (top panel) and $\sin(i_p)$ (bottom panel) for Gorchakov family members (small blue dots) identified by HCM analysis performed as part of this work using $\delta_c$$\,=\,$75~m~s$^{-1}$.  The proper elements for (5014) Gorchakov are marked with red triangles, while the proper elements for 238P are marked with yellow stars.
}
\label{figure:aei_gorchakov}
\end{figure}

The activity of 238P is strongly believed to be sublimation-driven based on numerical dust modeling of its activity in 2005 \citep{hsieh2009_238p} and observations of recurrent activity on two additional occasions in 2010 and 2016 \citep{hsieh2011_238p,hsieh2016_238p}, making the object a likely MBC.  The object's nucleus has been estimated to have $r_e$$\,\sim\,$0.4~km \citep[assuming $p_R$$\,=\,$0.05;][]{hsieh2009_238p}.  While we find that 238P is currently associated with the candidate Gorchakov family, \citet{haghighipour2009_mbcorigins} has suggested that it may have been a former member of the Themis family that has since migrated in eccentricity away from the family.  This hypothesis is supported by 238P's small $t_{ly}$ value (Table~\ref{table:aaproperties}), indicating that it is dynamically unstable, although the existence of a plausible dynamical pathway from the Themis family to 238P's current location has not yet been definitively demonstrated.


\subsubsection{The Lixiaohua Family}\label{section:lixiaohua}

\begin{figure}[htb!]
\centerline{\includegraphics[width=2.6in]{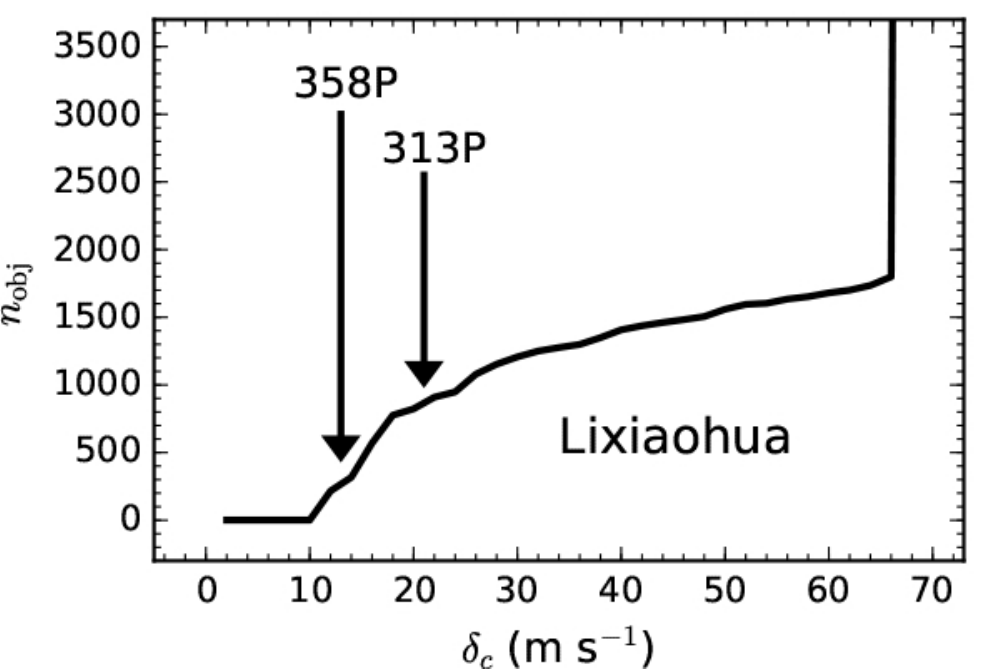}}
\caption{\small Plot of number of asteroids associated with (3556) Lixiaohua as a function of $\delta_c$, where the points at which 313P and 358P become linked with the family ($\delta_c$$\,=\,$21~m~s$^{-1}$ and $\delta_c$$\,=\,$13~m~s$^{-1}$, respectively) are marked with vertical arrows.
}
\label{figure:family_progression_lixiaohua}
\end{figure}

Active asteroids 313P/Gibbs (formerly designated P/2014 S4) and 358P/PANSTARRS (formerly designated P/2012 T1) have been previously linked to the Lixiaohua family \citep{hsieh2013_p2012t1,hsieh2015_313p}, which has been determined to be 155$\pm$36~Myr old \citep{novakovic2010_chaotictransport}.  The Lixiaohua family has a size-frequency distribution consistent with being the result of a catastrophic disruption event \citep{novakovic2010_lixiaohua,benavidez2012_sfds}.  313P becomes linked with the Lixiaohua family at $\delta_c$$\,=\,$21~m~s$^{-1}$, and 358P becomes linked with the family at $\delta_c$$\,=\,$13~m~s$^{-1}$ (Figure~\ref{figure:family_progression_lixiaohua}), both well within the optimum cut-off distance ($\delta_{c}$$\,=\,$45~m~s$^{-1}$) determined for the family by \citet{nesvorny2015_astfam_ast4}.

\begin{figure}[htb!]
\centerline{\includegraphics[width=2.1in]{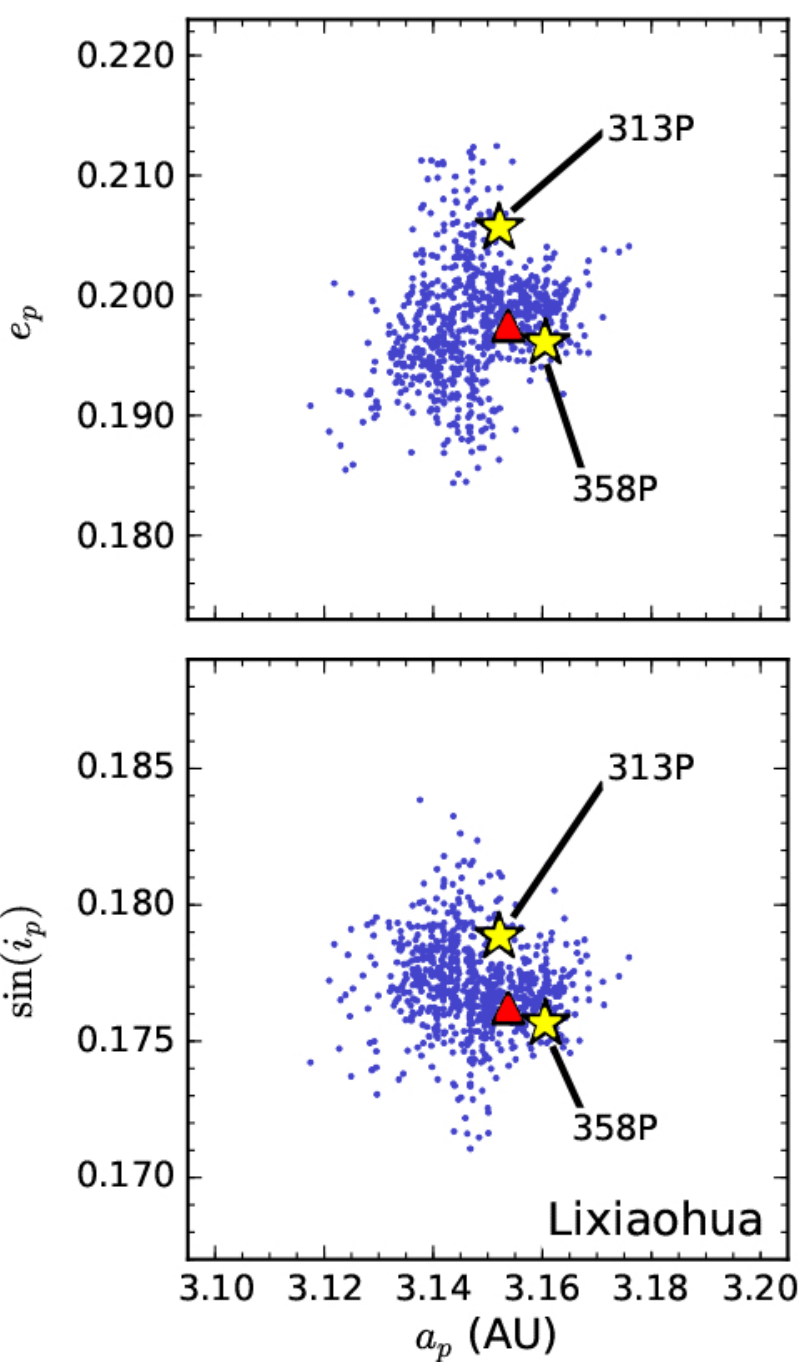}}
\caption{\small Plots of $a_p$ versus $e_p$ (top panel) and $\sin(i_p)$ (bottom panel) for Lixiaohua family members (small blue dots) identified by \citet{nesvorny2015_pdsastfam}.  The proper elements for (3556) Lixiaohua are marked with red triangles, while the proper elements for 313P and 358P are marked with yellow stars.
}
\label{figure:aei_lixiaohua}
\end{figure}

The family resides in a dynamically complex region of orbital element space in the asteroid belt (Figure~\ref{figure:aei_lixiaohua}), and is affected by several weak two- and three-body MMRs (cf.\ Figure~\ref{figure:aei_lixiaohua}) and potential close encounters with large asteroids, particularly Ceres, resulting in chaotic diffusion in all three proper elements ($a_p$, $e_p$, and $i_p$) \citep{novakovic2010_lixiaohua}.  Both 313P and 358P have small $t_{ly}$ values (Table~\ref{table:aaproperties}), indicating that they are relatively unstable over long timescales.  About 20\% of Lixiaohua family members also have similar or smaller $t_{ly}$ values, however, and so the small $t_{ly}$ values of 313P and 358P do not necessarily indicate that they are likely to be recently implanted interlopers, but may instead simply reflect the complex dynamical environment of the family resulting in general instability for a large number of its members.

The largest member of the family, (3556) Lixiaohua, has been spectroscopically classified as a C- or X-type asteroid \citep[cf.][]{nesvorny2005_spaceweathering}, and \changed{has been reported to have} $p_V$$\,=\,$0.035\changed{$\pm$0.004} and $r_e$$\,=\,$10.04\changed{$\pm$0.02}~km \citep{mainzer2016_neowise}.  Other family members have been classified as C-, D-, and X-type asteroids and \changed{have been reported to have} $\overline{p_V}$$\,=\,$0.044$\pm$0.009, indicating that they are likely to have primitive compositions.

The activity of 313P is strongly believed to be sublimation-driven based on both numerical dust modeling and observations showing that it has been active on at least two occasions in 2003 and 2014 \citep{hsieh2015_313p,jewitt2015_313p1,jewitt2015_313p2,hui2015_313p}, making the object a likely MBC.  Photometric monitoring of 358P while it was active in 2012 suggests that its activity is likely to be due to sublimation, making the object a likely MBC as well.


\subsubsection{The Mandragora Family}\label{section:mandragora}

\begin{figure}[htb!]
\centerline{\includegraphics[width=2.6in]{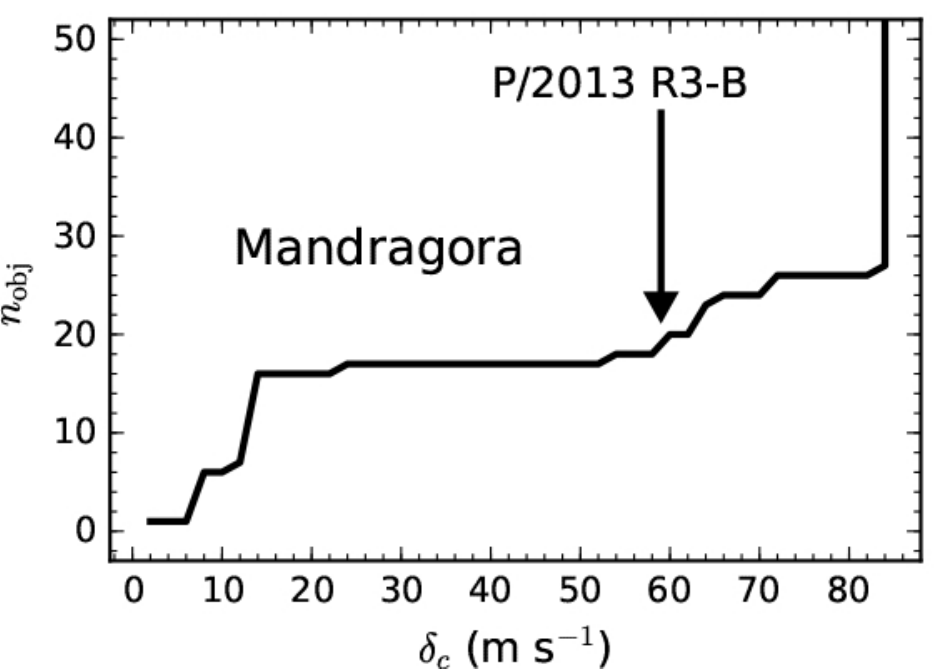}}
\caption{\small Plot of number of asteroids associated with (22280) Mandragora as a function of $\delta_c$, where the point at which P/2013 R3-B becomes linked with the family ($\delta_c$$\,=\,$59~m~s$^{-1}$) is marked with a vertical arrow.
}
\label{figure:family_progression_mandragora}
\end{figure}

We find that active asteroid component P/2013 R3-B (PANSTARRS) is linked to the recently identified Mandragora family, which has been determined to be 290$\pm$20~kyr old \citep{pravec2017_astclusters}.  P/2013 R3-B becomes linked with the Mandragora family at $\delta_c$$\,=\,$59~m~s$^{-1}$ (Figure~\ref{figure:family_progression_mandragora}), 
within the optimum cut-off distance ($\delta_{c}$$\,=\,$65~m~s$^{-1}$) determined for the family by \citet{pravec2017_astclusters}.  P/2013 R3-A is not formally linked to the family at the present time.

\begin{figure}[htb!]
\centerline{\includegraphics[width=2.1in]{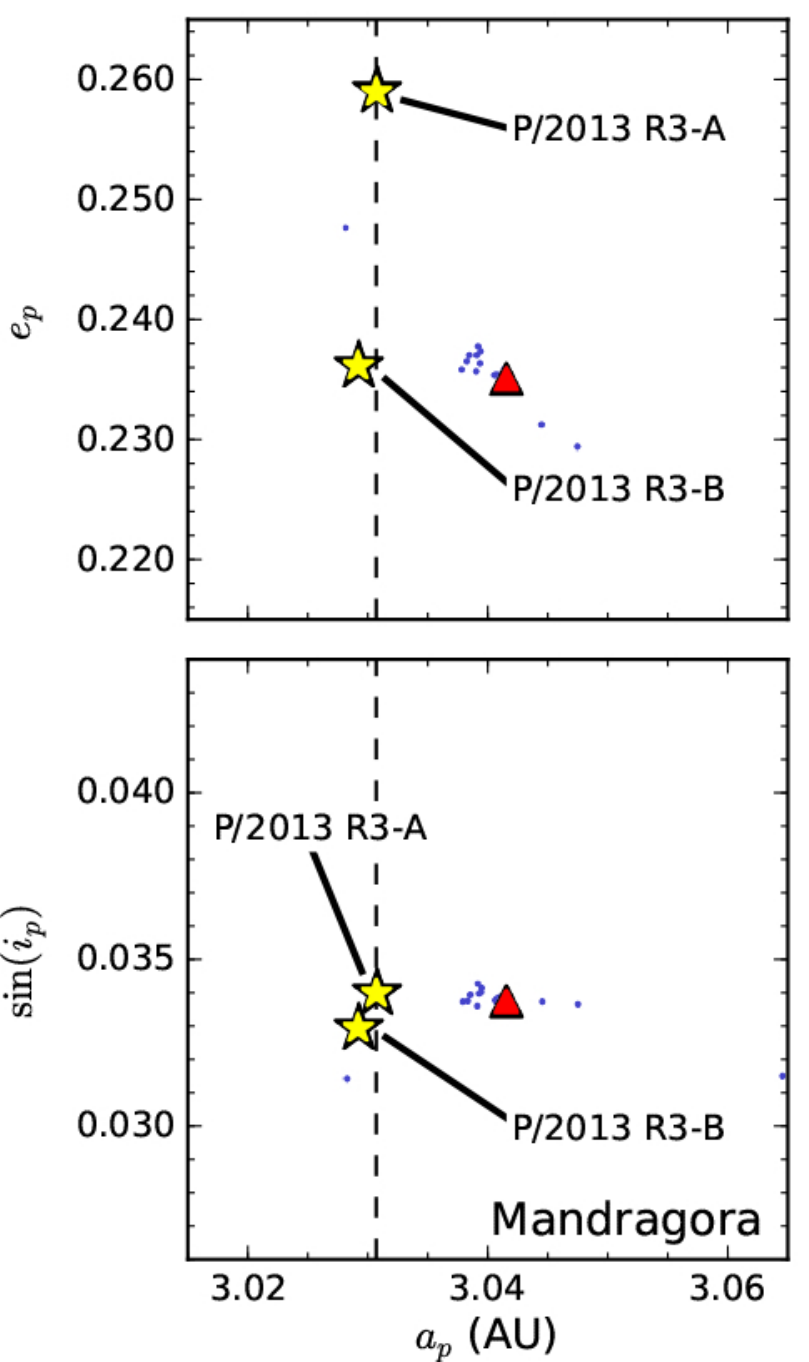}}
\caption{\small Plots of $a_p$ versus $e_p$ (top panel) and $\sin(i_p)$ (bottom panel) for candidate Mandragora family members (small blue dots) identified by \citet{pravec2017_astclusters}.  The proper elements for (22280) Mandragora are marked with red triangles, while the proper elements for P/2013 R3-A and P/2013 R3-B are marked with yellow stars.  Vertical dashed lines mark the semimajor axis position of the 9J:4A MMR at 3.0307~AU.
}
\label{figure:aei_mandragora}
\end{figure}

The 9J:4A MMR falls near the cluster, very nearly coinciding with the proper semimajor axes of P/2013 R3-A and P/2013 R3-B (Figure~\ref{figure:aei_mandragora}).  This means that those objects are likely to be dynamically unstable on long timescales, a conclusion supported by the objects' small $t_{ly}$ values (Table~\ref{table:aaproperties}), and may not in fact share a common origin with other Mandragora family members.  If the parent body of P/2013 R3-A and P/2013 R3-B was originally a member of the candidate Mandragora family though, destabilization by the 9J:4A MMR or non-gravitational outgassing forces could explain why P/2013 R3-A has diffused away from the family in proper eccentricity.  A third possibility is that P/2013 R3-A could simply not be linked to the Mandragora family due to poor proper element determination resulting from large uncertainties in the osculating orbital elements of both fragments.  A more detailed backward integration analysis like that performed by \citet{pravec2017_astclusters} for other members of the family would help to clarify the membership status of both fragments, and should be performed in the future.

The largest member of the Mandragora family, (22280) Mandragora, \changed{has been reported to have} $p_V$$\,=\,$0.046\changed{$\pm$0.006} and $r_e$$\,=\,$4.9\changed{$\pm$0.1}~km \citep{mainzer2016_neowise,pravec2017_astclusters}.  No family members have been taxonomically classified, but those with measured albedos \changed{have been reported to have} $\overline{p_V}$$\,=\,$0.056\changed{$\pm$0.019}, indicating that they are likely to have primitive compositions.  This conclusion is also supported by the C-type-like $V-R$ colors measured for the two largest members of the family reported by \citet{pravec2017_astclusters}.

When P/2013 R3 was discovered, the object had already split into multiple fragments, which then disintegrated further over the following several months.  Analysis of follow-up observations suggested that the comet likely broke apart due to stresses from rapid rotation \citep{jewitt2014_p2013r3,jewitt2017_p2013r3}.  \citet{jewitt2014_p2013r3} concluded that gas pressure alone was insufficient for causing the catastrophic disruption of the comet, although individual fragments were observed to exhibit secondary dust emission behavior indicative of being driven by sublimation, perhaps of newly exposed interior ices, making the object a likely MBC.


\subsubsection{The Themis and Beagle Families}\label{section:themis_beagle}

\begin{figure}[htb!]
\centerline{\includegraphics[width=2.6in]{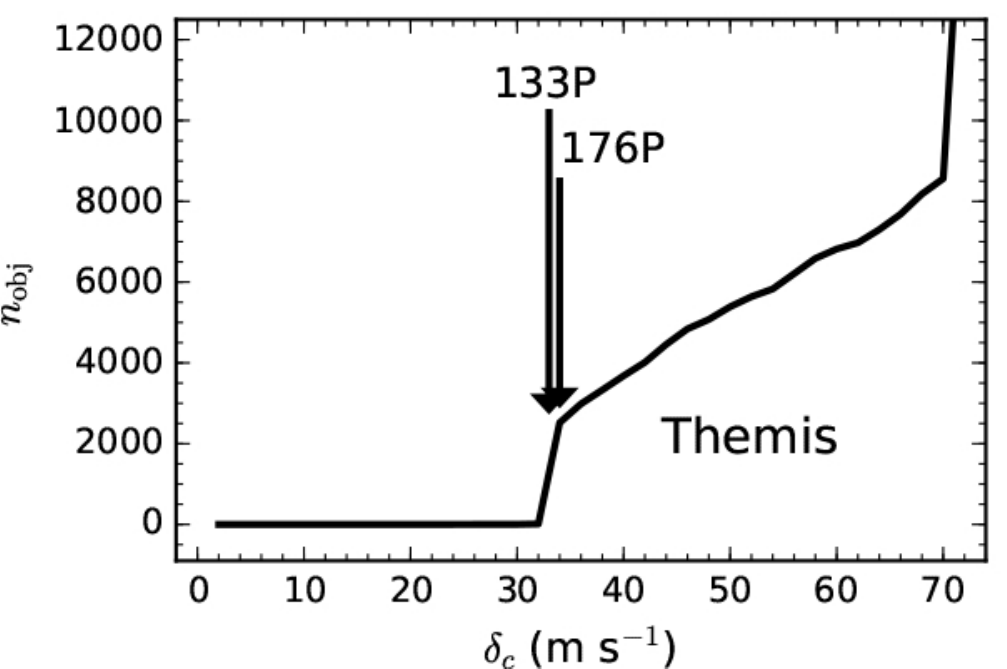}}
\caption{\small Plot of number of asteroids associated with (24) Themis as a function of $\delta_c$, where the points at which 133P and 176P become linked with the family ($\delta_c$$\,=\,$33~m~s$^{-1}$ and $\delta_c$$\,=\,$34~m~s$^{-1}$, respectively) are marked with vertical arrows.
}
\label{figure:family_progression_themis}
\end{figure}

Active asteroids 133P (also designated (7968) Elst-Pizarro) and 176P (also designated (118401) LINEAR) have been previously linked to the Themis family \citep[e.g.,][]{boehnhardt1998_133p,hsieh2009_htp}.  While MBC 288P was previously found to be associated with the Themis family, we find that it becomes linked with the family at $\delta_c$$\,=\,$77~m~s$^{-1}$, outside the nominally established $\delta_c$ for the family. 133P becomes linked with the family at $\delta_c$$\,=\,$33~m~s$^{-1}$ and 176P becomes linked with the family at $\delta_c$$\,=\,$34~m~s$^{-1}$ (Figure~\ref{figure:family_progression_themis}), both well within the optimum cut-off distance ($\delta_{c}$$\,=\,$60~m~s$^{-1}$) determined for the family by \citet{nesvorny2015_astfam_ast4}.  \citet{nesvorny2003_dustbands} estimated the age of the Themis family to be 2.5$\pm$1.0~Gyr, but due to large uncertainties caused by its old age and dynamical environment, other estimates for the age of the Themis family range from as little as 500~Myr to nearly the age of the solar system ($\sim$4.5~Gyr) \citep{spoto2015_astfamages,carruba2016_oldestfamilies}.

\begin{figure}[htb!]
\centerline{\includegraphics[width=2.6in]{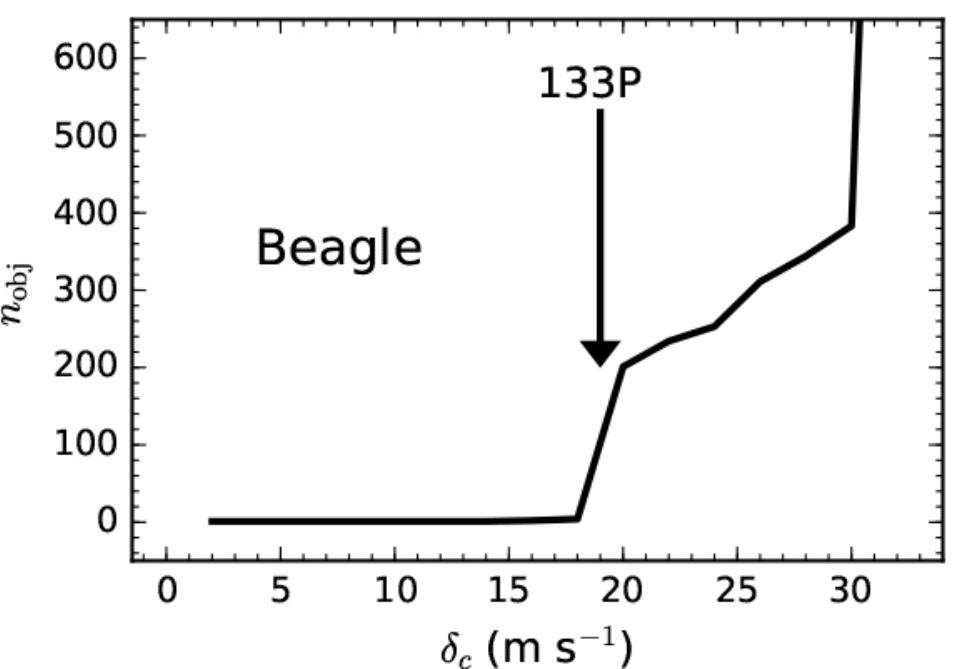}}
\caption{\small Plot of number of asteroids associated with (656) Beagle as a function of $\delta_c$, where the point at which 133P becomes linked with the family ($\delta_c$$\,=\,$19~m~s$^{-1}$) is marked with a vertical arrow.
}
\label{figure:family_progression_beagle}
\end{figure}

Meanwhile, 133P has been previously determined to also be linked to the Beagle family, which is a sub-family of the Themis family and has been estimated to be $<\,$10~Myr old \citep{nesvorny2008_beagle}.  133P becomes linked with the Beagle family at $\delta_c$$\,=\,$19~m~s$^{-1}$ (Figure~\ref{figure:family_progression_beagle}), 
within the optimum cut-off distance ($\delta_{c}$$\,=\,$25~m~s$^{-1}$) determined for the family by \citet{nesvorny2015_astfam_ast4}.

\begin{figure}[htb!]
\centerline{\includegraphics[width=2.1in]{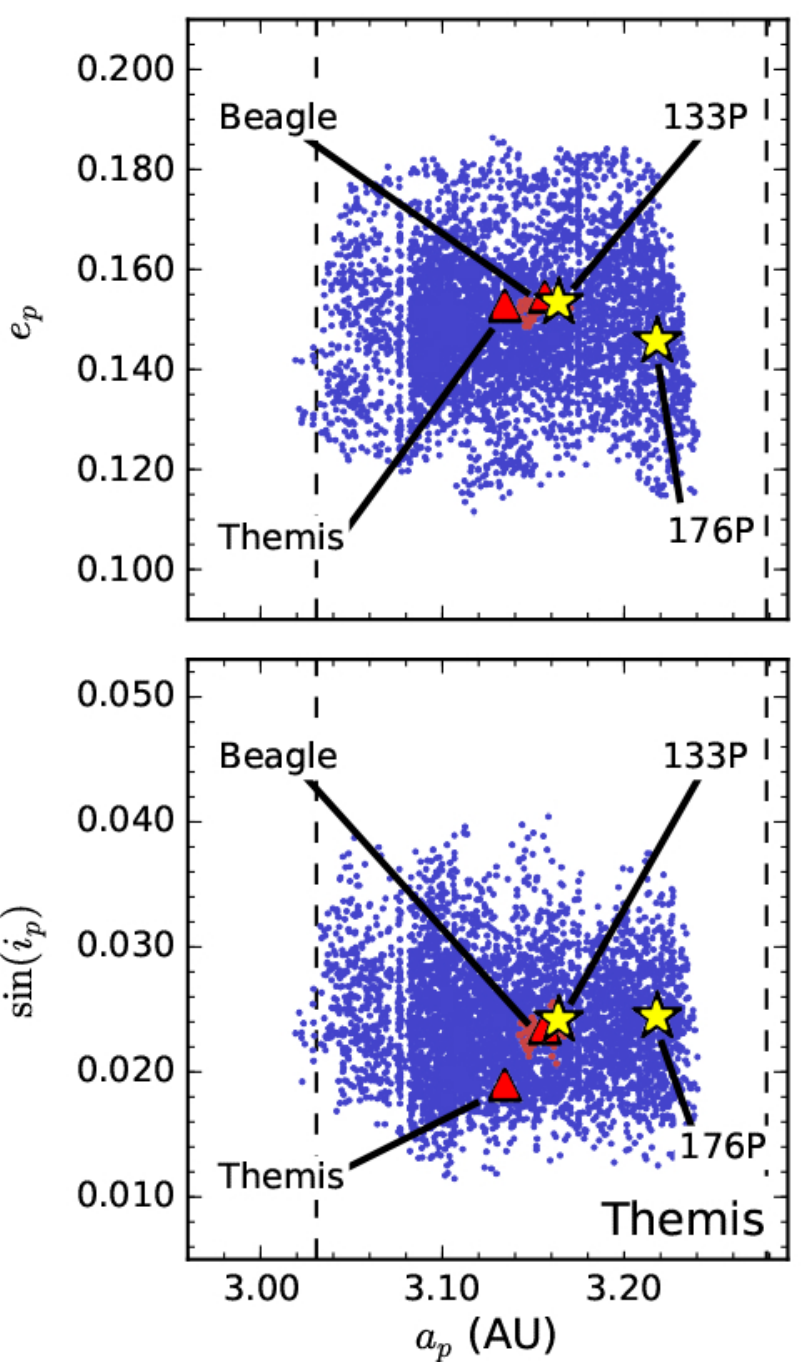}}
\caption{\small Plots of $a_p$ versus $e_p$ (top panel) and $\sin(i_p)$ (bottom panel) for Themis family members (small blue dots; using $\delta_c$$\,=\,$60~m~s$^{-1}$) and Beagle family members (small pale red dots) identified by \citet{nesvorny2015_pdsastfam}.  The proper elements for (24) Themis and (656) Beagle are marked with red triangles, while the proper elements for 133P and 176P are marked with yellow stars.
Vertical dashed lines mark the semimajor axis positions of the 9J:4A (left) and 2J:1A (right) MMRs
at 3.0307~AU and 3.2783~AU, respectively.
}
\label{figure:aei_themis}
\end{figure}

\begin{figure}[htb!]
\centerline{\includegraphics[width=2.1in]{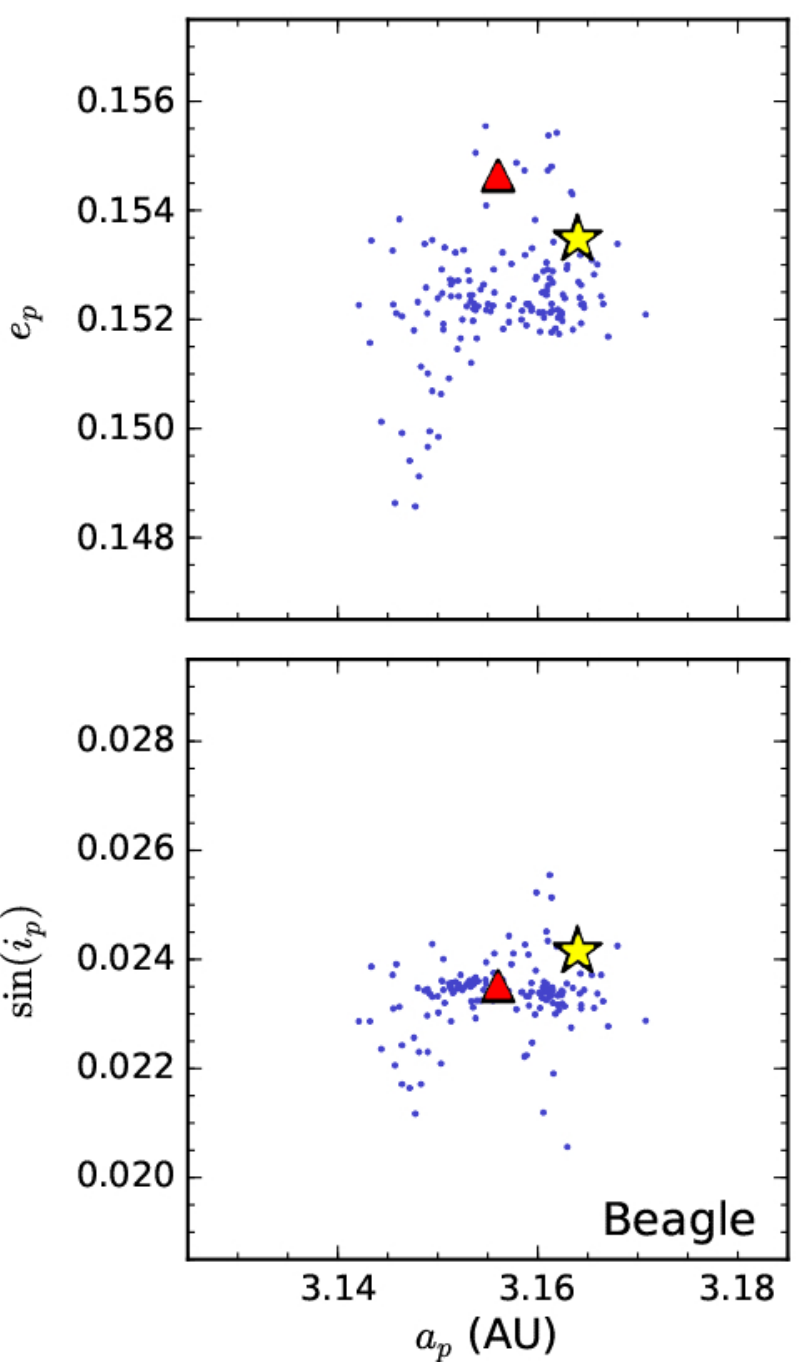}}
\caption{\small Plots of $a_p$ versus $e_p$ (top panel) and $\sin(i_p)$ (bottom panel) for the Beagle family (small blue dots; using $\delta_c$$\,=\,$25~m~s$^{-1}$).  The proper elements for (656) Beagle are marked with red triangles, while the proper elements for 133P are marked with yellow stars.
}
\label{figure:aei_beagle}
\end{figure}

The Themis family was one of the original asteroid families identified by \citet{hirayama1918_astfam}.  Due to the fact that the Themis family is adjacent to the 2J:1A MMR (cf.\ Figure~\ref{figure:aei_themis}), it is believed that a significant number of family members may have been captured and scattered by the resonance since the family's formation \citep{morbidelli1995_familyresonances}.

The largest member of the family, (24) Themis, has been spectroscopically classified as a B- or C-type asteroid \citep{neese2010_taxonomy}, and \changed{has been reported to have} $p_V$$\,=\,$0.069\changed{$\pm$0.010}, $r_e$$\,=\,$97.8\changed{$\pm$2.2}~km, and an estimated density of $\rho$$\,=\,$1.81$\pm$0.67~kg~m$^{-3}$ \citep{mainzer2016_neowise,carry2012_astdensities}.  Notably, a near-infrared  absorption feature attributed to water ice frost was detected in spectra of both Themis \citep{rivkin2010_themis,campins2010_themis} and another large member of the family, (90) Antiope \citep{hargrove2015_antiope}. No evidence of outgassing in the form of spectroscopic detections of outgassing has yet been found, though \citep{jewitt2012_themiscybele,mckay2017_themisceres}.  The Themis family in general is dominated by C-complex asteroids, many of which exhibit spectra indicative of aqueously altered mineralogy and are similar to carbonaceous chondrite meteorites \citep{florczak1999_themisspectra,ziffer2011_themisveritas}.

The Beagle family is entirely contained within the Themis family (cf.\ Figure~\ref{figure:aei_themis}), suggesting that it formed from the fragmentation of a parent body that was itself a Themis family member.  Despite the now commonly-used name of the family, \citet{nesvorny2008_beagle} noted that it is possible that (656) Beagle may not actually be a real member of the family that bears its name, as its slight eccentricity offset from the other family members (Figure~\ref{figure:aei_beagle}) would require the invocation of an unusual ejection velocity field to explain.

No formal taxonomic classification has been reported for the largest member of the family, (656) Beagle, but it has a spectrum consistent with C-complex asteroids \citep{kaluna2016_spaceweathering,fornasier2016_themisbeagle}, and \changed{has been reported to have} $p_V$$\,=\,$0.045\changed{$\pm$0.005} and $r_e$$\,=\,$31.3\changed{$\pm$0.3}~km \citep{mainzer2016_neowise}.  Other family members have been taxonomically classified as B- and C-type asteroids, and \changed{have been reported to have} $\overline{p_V}$$\,=\,$0.080$\pm$0.014 (cf.\ Table~\ref{table:family_associations}), indicating that they are likely to have primitive compositions.

Based on the simultaneous detection of some objects with phyllosilicate absorption features and the presence of the apparently ice-bearing 133P in the family, \citet{kaluna2016_spaceweathering} concluded that the Beagle parent body was most likely composed of a heterogeneous mixture of ice and aqueously altered material.  Meanwhile, spectroscopic observations of samples of both Beagle and Themis family asteroids showed that Beagle family asteroids are spectrally bluer, have higher albedos, and exhibit smaller spectral slope variability than background Themis family asteroids, suggesting that the Beagle parent body could have been a particularly blue and bright interior fragment of the original Themis parent body \citep{fornasier2016_themisbeagle}.

133P has been observed to be active during four perihelion passages (in 1996, 2002, 2007, and 2013) with intervening periods of inactivity, where dust modeling results indicate that dust emission took place over periods of months during its 1996, 2002, and 2013 active epochs \citep{boehnhardt1998_133p,hsieh2004_133p,jewitt2014_133p}.
As such, 133P's activity is strongly believed to be the result of sublimation of volatile ices, although it is possible that the object's rapid rotation ($P_{\rm rot}$$\,=\,$3.471$\pm$0.001~hr) may also play a role in helping to eject dust particles \citep{hsieh2004_133p}.  133P's nucleus has been taxonomically classified as a B- or F-type asteroid \citep{bagnulo2010_133p,licandro2011_133p176p} and \changed{has been reported to have} $p_R$$\,=\,$0.05$\pm$0.02 and $r_e$$\,=\,$1.9$\pm$0.3~km \citep{hsieh2009_albedos}.  While the presence of water ice has not yet been definitively spectroscopically confirmed on 133P, \citet{rousselot2011_133p} reported that its spectrum could be consistent with a mixture of water ice, black carbon, tholins, and silicates, but acknowledged that such a compositional interpretation was not unique.

176P's nucleus has been taxonomically classified as a B-type asteroid \citep{licandro2011_133p176p} and \changed{has been reported to have} $p_R$$\,=\,$0.06$\pm$0.02 and $r_e$$\,=\,$2.0$\pm$0.2~km \citep{hsieh2009_albedos}.  Numerical dust modeling of its activity in 2005 indicated that it was likely to be due to a prolonged dust emission event, pointing to sublimation as the mostly likely driver of the activity \citep{hsieh2011_176p}, making the object a likely MBC.  However, numerous observations during the object's next perihelion passage in 2011 revealed no evidence of recurrent activity, which could either suggest that the object did not actually exhibit sublimation-driven activity when observed in 2005, or that the object's activity had simply become attenuated to an undetectable level during the following orbit passage \citep{hsieh2014_176p}.


\subsubsection{The Theobalda Family}\label{section:theobalda}

\begin{figure}[htb!]
\centerline{\includegraphics[width=2.6in]{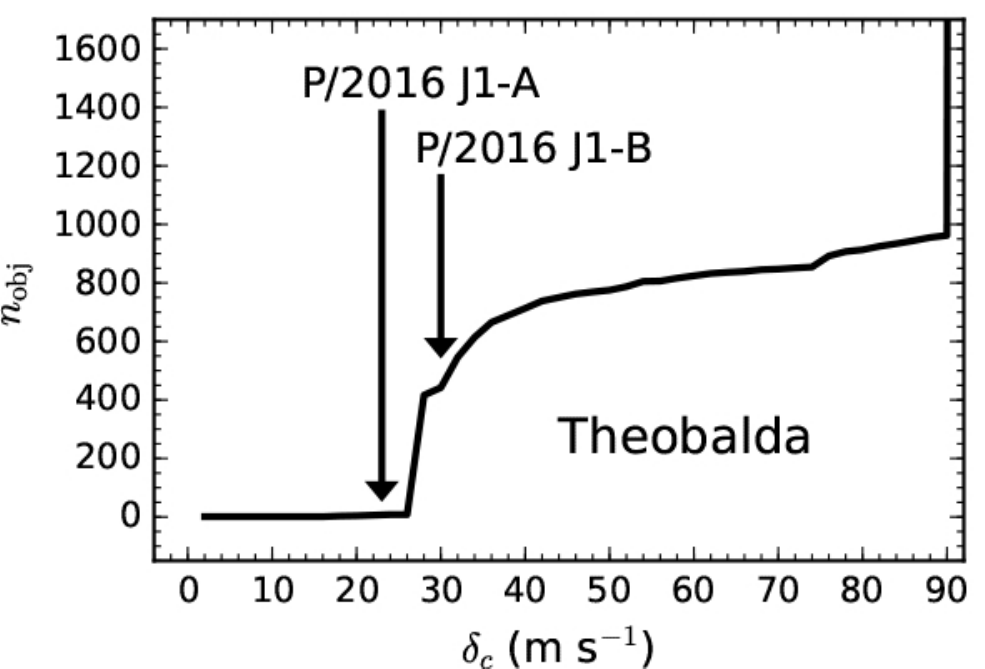}}
\caption{\small Plot of number of asteroids associated with (778) Theobalda as a function of $\delta_c$, where the points at which P/2016 J1-A and P/2016 J1-B become linked with the family ($\delta_c$$\,=\,$23~m~s$^{-1}$ and $\delta_c$$\,=\,$30~m~s$^{-1}$, respectively) are marked with vertical arrows.
}
\label{figure:family_progression_theobalda}
\end{figure}

\begin{figure}[htb!]
\centerline{\includegraphics[width=2.1in]{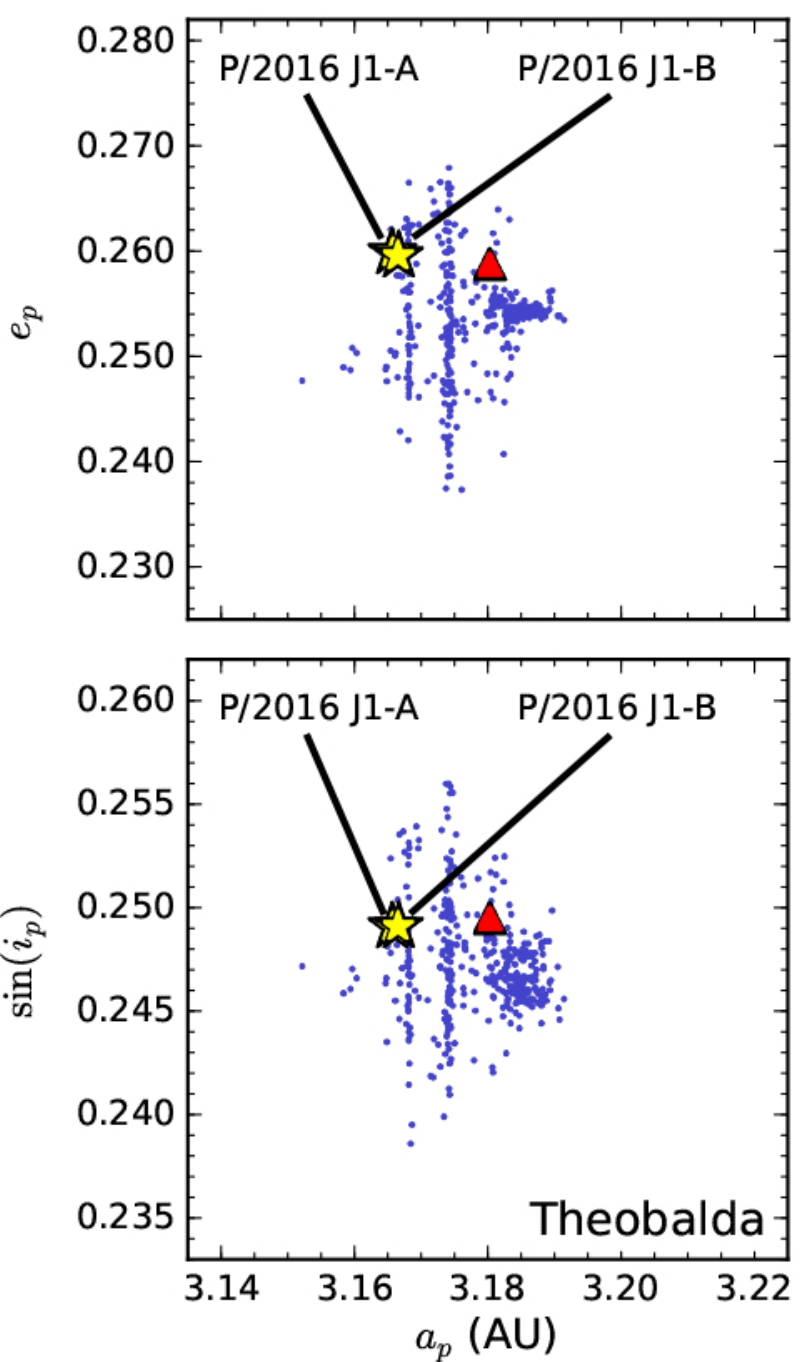}}
\caption{\small Plots of $a_p$ versus $e_p$ (top panel) and $\sin(i_p)$ (bottom panel) for Theobalda family members (small blue dots) identified by \citet{nesvorny2015_astfam_ast4}.  The proper elements for (778) Theobalda are marked with red triangles, while the proper elements for P/2016 J1-A and P/2016 J1-B are marked with yellow stars.
}
\label{figure:aei_theobalda}
\end{figure}

We find that active asteroid P/2016 J1-A/B (PANSTARRS) is linked to the Theobalda family.  P/2016 J1-A becomes linked with the Theobalda family at $\delta_c$$\,=\,$23~m~s$^{-1}$, and P/2016 J1-B becomes linked with the family at $\delta_c$$\,=\,$30~m~s$^{-1}$ (Figure~\ref{figure:family_progression_theobalda}), both well within the optimum cut-off distance ($\delta_{c}$$\,=\,$85~m~s$^{-1}$) determined for the family by \citet{novakovic2010_theobalda}.  \citet{novakovic2010_theobalda} further determined via two independent methods (chaotic chronology and backward integration) that the family was likely produced 6.9$\pm$2.3~Myr ago by a cratering impact on a $d$$\,=\,$78$\pm$9~km parent body.
A detailed dynamical analysis of this family was performed by \citet{novakovic2010_theobalda}, who found that it is crossed by several three-body MMRs (Figure~\ref{figure:aei_theobalda}) making the region significantly chaotic.

The largest member of the family, (778) Theobalda, has been spectroscopically classified as a F-type asteroid \citep{neese2010_taxonomy}, and \changed{has been reported to have} $p_V$$\,=\,$0.079\changed{$\pm$0.010} and $r_e$$\,=\,$27.7\changed{$\pm$0.4}~km \citep{mainzer2016_neowise}.  Other family members have been taxonomically classified as C-, F-, and X-type asteroids and \changed{have been reported to have} $\overline{p_V}$$\,=\,$0.062$\pm$0.016 (cf.\ Table~\ref{table:family_associations}), indicating that they are likely to have primitive compositions.

P/2016 J1 was characterized by \citet{hui2017_p2016j1} and \citet{moreno2017_p2016j1}, who found mass loss rates of $\lesssim\,$1~kg~s$^{-1}$ for both components of the object (P/2016 J1-A and P/2016 J1-B).  Both sets of authors also found that both components were continuously active over a period of three to nine months, strongly suggesting that the activity was sublimation-driven, making the object a likely MBC. \citet{hui2017_p2016j1} also estimated that the two largest fragments, J1-A and J1-B, have radii of 140$\,$m$\,<\,$$r_{e}$$\,<\,$900$\,$m and 40$\,$m$\,<\,$$r_{e}$$\,<\,$400$\,$m, respectively, and broadband colors similar to C- or G-type asteroids.


\subsubsection{The 288P Family}\label{section:288P}

\begin{figure}[htb!]
\centerline{\includegraphics[width=2.1in]{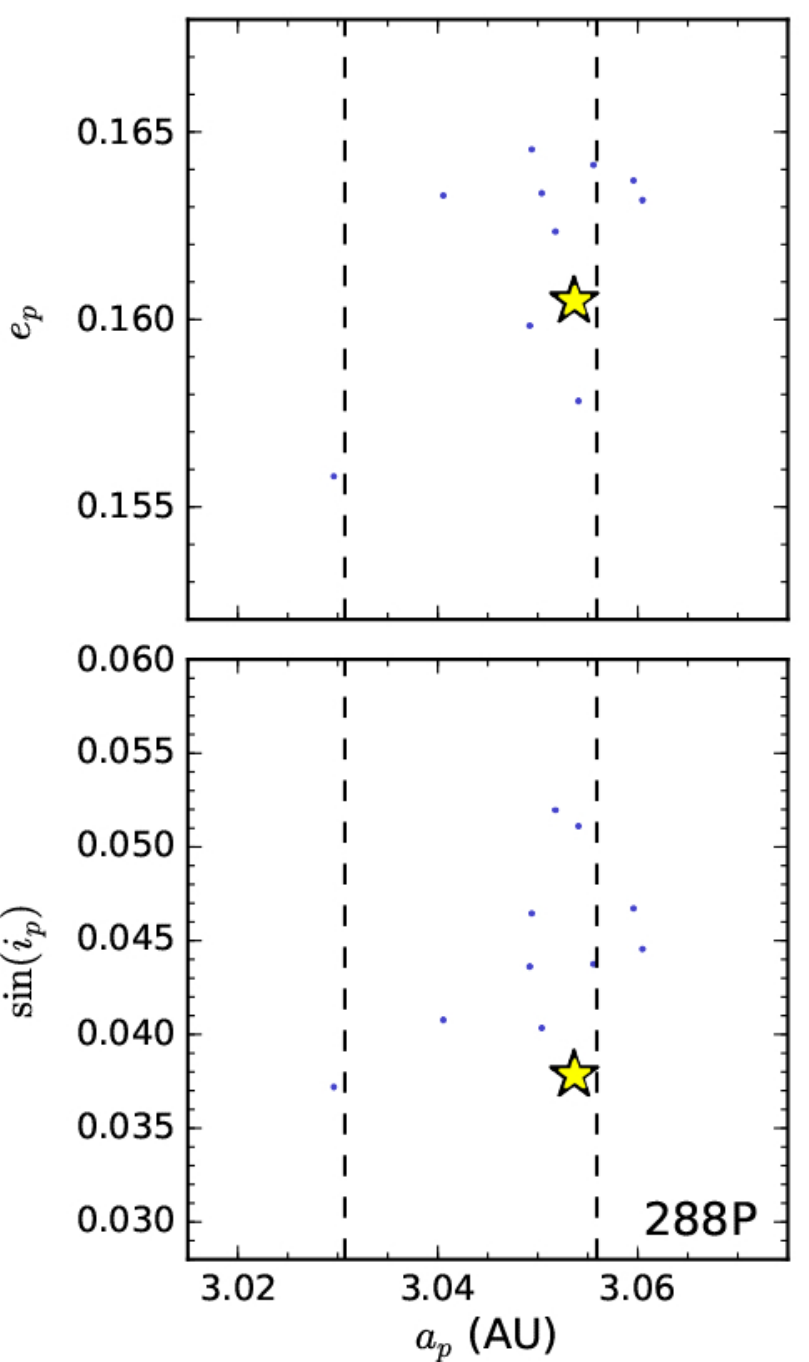}}
\caption{\small Plots of $a_p$ versus $e_p$ (top panel) and $\sin(i_p)$ (bottom panel) for 288P family members (small blue dots) identified by \citet{novakovic2012_288p}.  The proper elements for 288P are marked with yellow stars.  Vertical dashed lines mark the semimajor axis positions of the 9J:4A (left) and 20J:9A (right) MMRs at 3.0307~AU and 3.0559~AU, respectively.
}
\label{figure:aei_288p}
\end{figure}

Active asteroid 288P/(300163) 2006 VW$_{139}$ has been previously linked to a 7.5$\pm$0.3~Myr-old asteroid family designated as the 288P family \citep{novakovic2012_288p}.
The 11-member 288P family was analyzed in detail by \citet{novakovic2012_288p} who found that it was likely formed in a disruptive event characterized as being intermediate between a catastrophic disruption and a cratering event.  It is located in close proximity to the Themis family, with which it merges at $\delta_c$$\,\sim\,$75~m~s$^{-1}$, separated by a number of weak two- and three-body MMRs.  These MMRs may contribute to a number of dynamically unstable interlopers, which were excluded from the family by \citet{novakovic2012_288p} based on a backward integration method (BIM) analysis \citep[cf.][]{nesvorny2002_karin}.  The 288P family is roughly bound by the 9J:4A MMR at 3.0307~AU on one side and is crossed by the 20J:9A MMR at 3.0559~AU (Figure~\ref{figure:aei_288p}).

\begin{figure}[htb!]
\centerline{\includegraphics[width=2.6in]{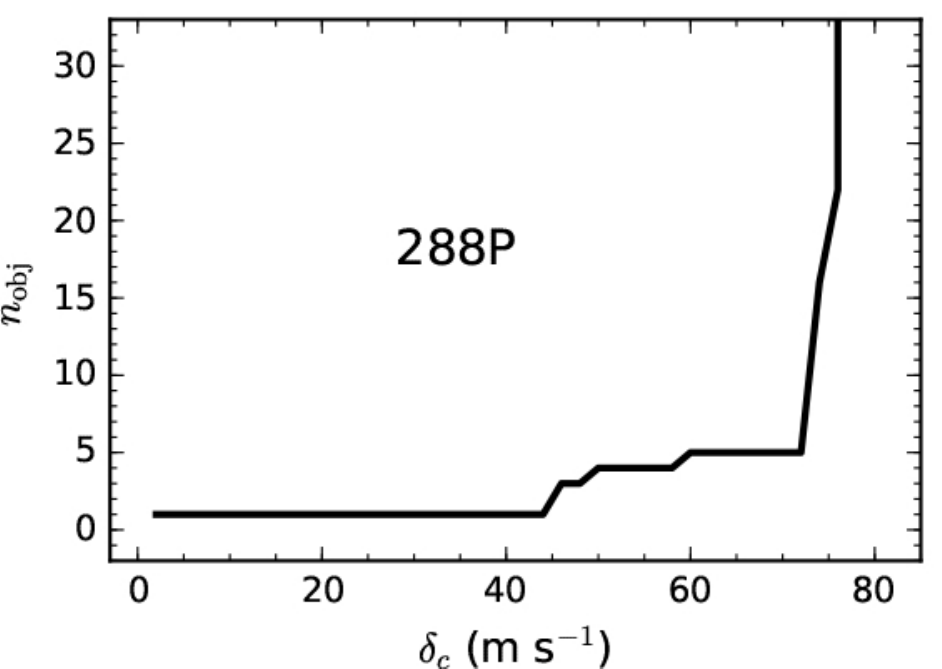}}
\caption{\small Plot of number of asteroids associated with 288P as a function of $\delta_c$.
}
\label{figure:family_progression_288p}
\end{figure}

The two members of the family which have had their albedos measured \changed{have been reported to have} ${p_V}$$\,=\,$0.077$\pm$0.037 and ${p_V}$$\,=\,$0.103$\pm$0.077, giving  $\overline{p_V}$$\,=\,$0.090$\pm$0.021 (cf.\ Table~\ref{table:family_associations}), indicating that they may have relatively primitive compositions.
The activity of 288P seen in 2011 is believed to be sublimation-driven based on numerical dust modeling \citep{hsieh2012_288p,agarwal2016_288p}, making the object a likely MBC, where this conclusion was further strengthened by the recent confirmation in 2016 that the object is recurrently active \citep{agarwal2016_288p}.  The nucleus of 288P has been classified as a C-type asteroid, has an effective absolute magnitude of $H_V$$\,=\,$17.0$\pm$0.1 (equivalent to $r_e$$\,\sim\,$1.3~km, assuming $p_V$$\,=\,$0.04), and has recently been confirmed to be a binary system with approximately equally sized components \citep{licandro2013_288p,agarwal2016_288p,agarwal2017_288p}.

\subsection{Disrupted Asteroid Family Associations}\label{section:dafamilies}


\subsubsection{The Adeona Family}\label{section:adeona}

\begin{figure}[htb!]
\centerline{\includegraphics[width=2.6in]{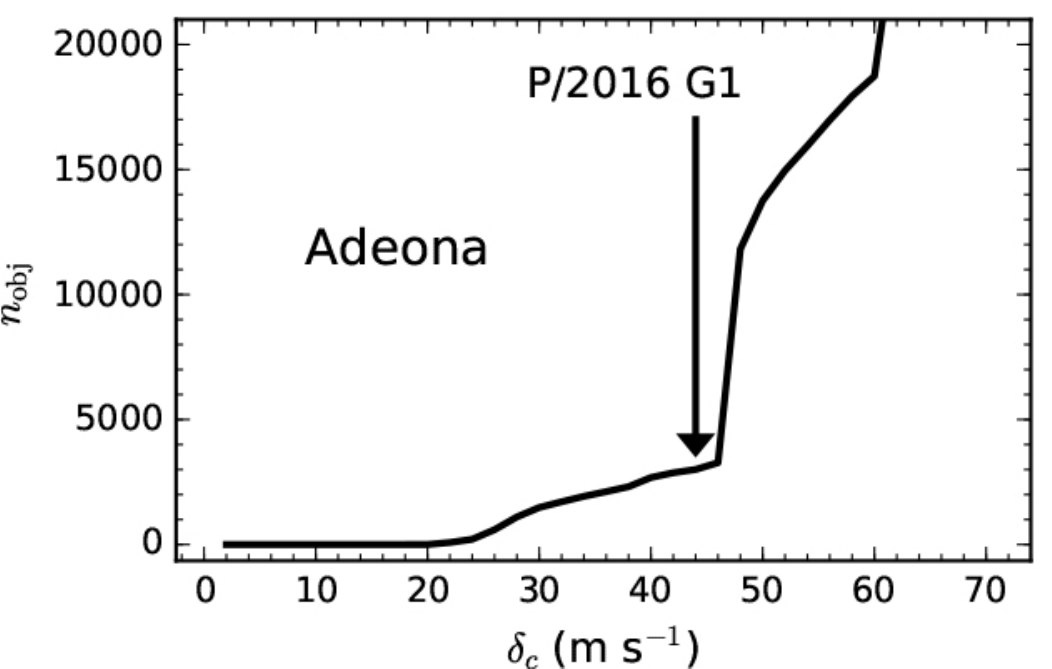}}
\caption{\small Plot of number of asteroids associated with (145) Adeona as a function of $\delta_c$, where the point at which P/2016 G1 becomes linked with the family ($\delta_c$$\,=\,$44~m~s$^{-1}$) is marked with a vertical arrow.
}
\label{figure:family_progression_adeona}
\end{figure}

We find that active asteroid P/2016 G1 (PANSTARRS) is linked to the Adeona family, which is estimated to have formed in a cratering event 
$\sim$700~Myr ago \citep{benavidez2012_sfds,carruba2016_ejectionfields,milani2017_astfamages}.  P/2016 G1 becomes linked with the family at $\delta_c$$\,=\,$44~m~s$^{-1}$ (Figure~\ref{figure:family_progression_adeona}), within the optimum cut-off distance ($\delta_{c}$$\,=\,$50~m~s$^{-1}$) determined for the family by \citet{nesvorny2015_astfam_ast4}.

\begin{figure}[htb!]
\centerline{\includegraphics[width=2.1in]{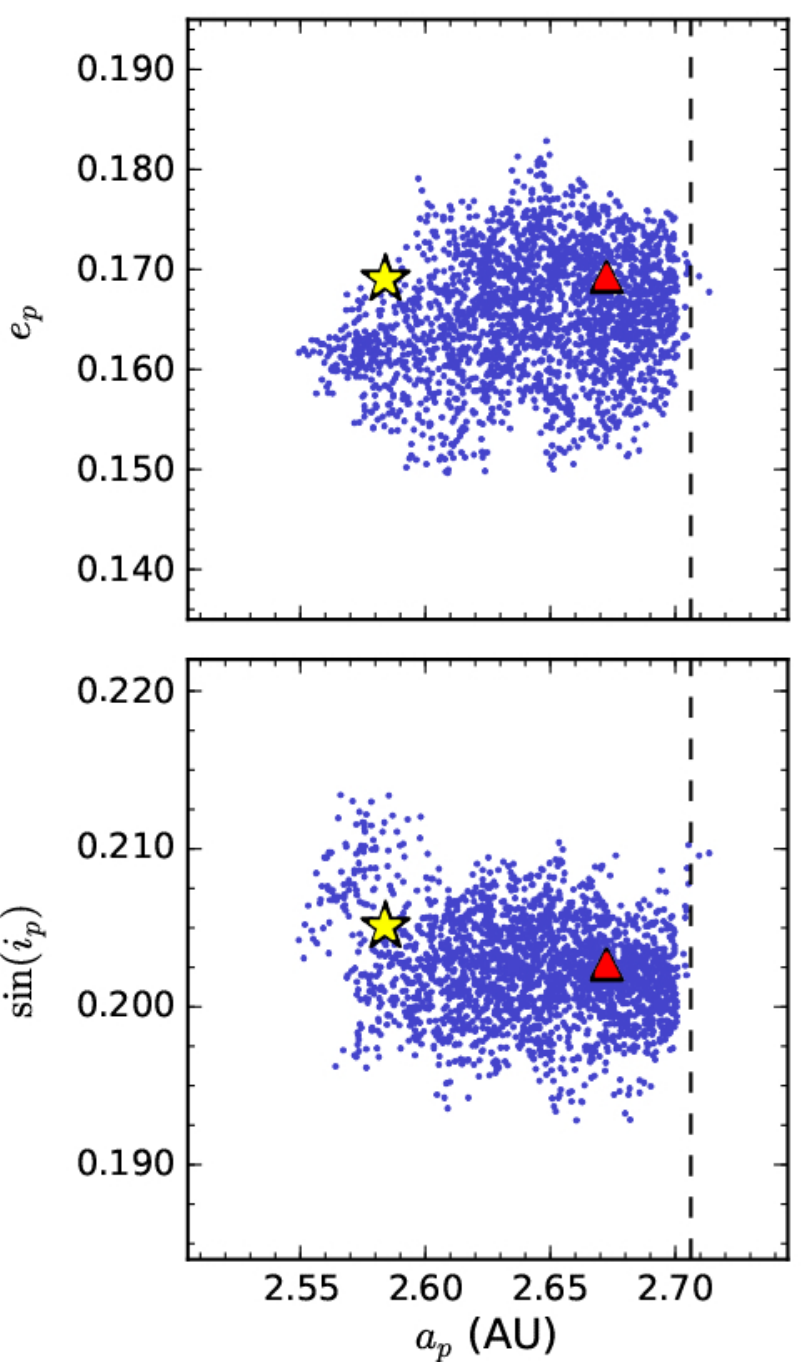}}
\caption{\small Plots of $a_p$ versus $e_p$ (top panel) and $\sin(i_p)$ (bottom panel) for Adeona family members (small blue dots) identified by \citet{nesvorny2015_pdsastfam}.  The proper elements for (145) Adeona are marked with red triangles, while the proper elements for P/2016 G1 are marked with yellow stars.  Vertical dashed lines mark the semimajor axis position of the 8J:3A MMR at 2.7062~AU.
}
\label{figure:aei_adeona}
\end{figure}

The Adeona family's orbital evolution was investigated in detail by \citet{carruba2003_gefionadeona}, who found that perturbations from large asteroids like Ceres could impact the inferred ejection velocities of family members, but should have minimal effect on the spread of the family's semimajor axis distribution.  The sharp cut-off of the family at the 8J:3A MMR at 2.706~AU (Figure~\ref{figure:aei_adeona}) is attributed to family members drifting into the resonance under the influence of the Yarkovsky effect and becoming scattered in eccentricity, thus becoming unrecognizable as family members.  The family is also crossed by several other two-body, three-body, and secular resonances \citep[][]{carruba2003_gefionadeona}. 

The Adeona family is notable in that the members of the family that have been spectroscopically classified have C- and Ch-type classifications, while the nearby background population is dominated by S-type asteroids.  The largest member of the family, (145) Adeona, has been spectroscopically classified as a C- or Ch-type asteroid \citep{neese2010_taxonomy}, and \changed{has been reported to have} $p_V$$\,=\,$0.061\changed{$\pm$0.010}, $r_e$$\,=\,$63.9\changed{$\pm$0.2}~km, and $\rho$$\,=\,$1.18$\pm$0.34~kg~m$^{-3}$ \citep{mainzer2016_neowise,carry2012_astdensities}.  Adeona has been spectroscopically characterized by \citet{busarev2015_asteroidspectra} who also found spectroscopic features indicative of hydrated silicates and hydrated oxides.  An unexplained sharp increase in reflectivity between 0.4~$\mu$m and 0.7~$\mu$m was also noted, and interpreted as possibly being indicative of a cloud of sublimed or frozen ice particles, but this interpretation has yet to be confirmed.  Other family members \changed{have been reported to have} $\overline{p_V}$$\,=\,$0.060$\pm$0.011 (cf.\ Table~\ref{table:family_associations}), indicating that they are likely to have primitive compositions.  The family's C-type members also exhibit evidence of aqueous alteration, and have been judged to be consistent with the breakup of a CM chondrite-like body \citep{mothediniz2005_familyspectroscopy}.

\citet{moreno2016_p2016g1} found that P/2016 G1's active behavior is best interpreted as the result of a short duration event about one year prior to perihelion, consistent with an impact which then led to the observed disintegration of the object, making the object a likely disrupted asteroid.
Those authors also found an upper limit of $r_e$$\,\sim\,$50~m for any post-disruption fragments.


\subsubsection{The Baptistina Family}\label{section:baptistina}

\begin{figure}[htb!]
\centerline{\includegraphics[width=2.6in]{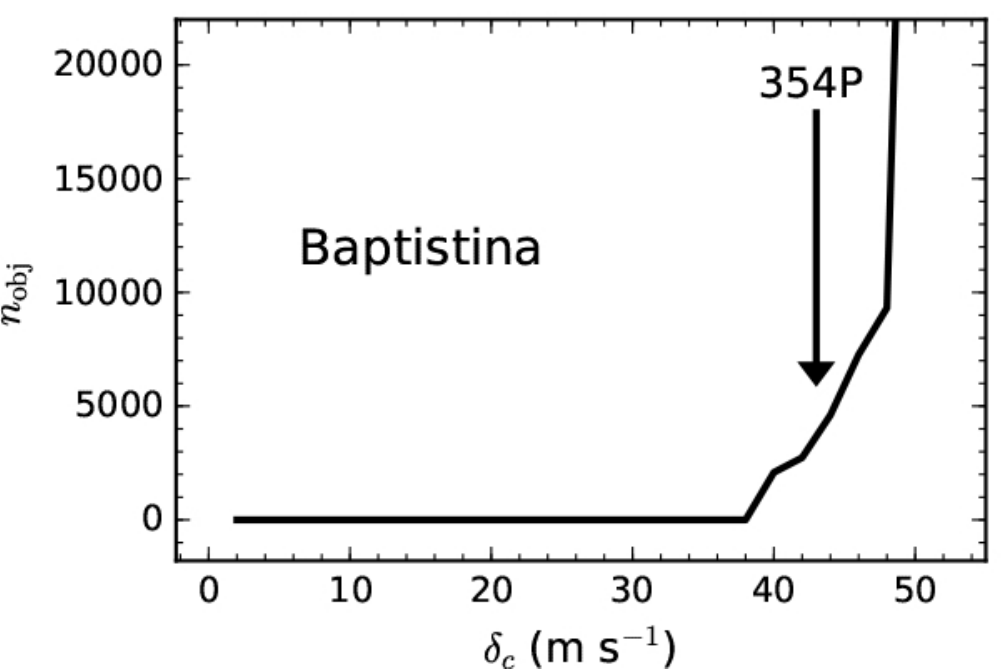}}
\caption{\small Plot of number of asteroids associated with (298) Baptistina as a function of $\delta_c$, where the point at which 354P becomes linked with the family ($\delta_c$$\,=\,$43~m~s$^{-1}$) is marked with a vertical arrow.
}
\label{figure:family_progression_baptistina}
\end{figure}

We find that disrupted asteroid 354P/LINEAR (formerly designated P/2010 A2) is linked to the Baptistina family \citep[despite being initially suspected of being a member of the Flora family; e.g.,][]{snodgrass2010_p2010a2}.  \citet{masiero2012_baptistina} determined the family's age to be between 140 to 320 Myr old, depending on the physical properties assumed for its family members, while \citet{carruba2016_ejectionfields} estimated the family's age to be 110$\pm$10~Myr old.  354P becomes linked with the family at $\delta_c$$\,=\,$43~m~s$^{-1}$ (Figure~\ref{figure:family_progression_baptistina}), within the optimum cut-off distance ($\delta_{c}$$\,=\,$48~m~s$^{-1}$) determined for the family by \citet{nesvorny2015_astfam_ast4}.

\begin{figure}[htb!]
\centerline{\includegraphics[width=2.1in]{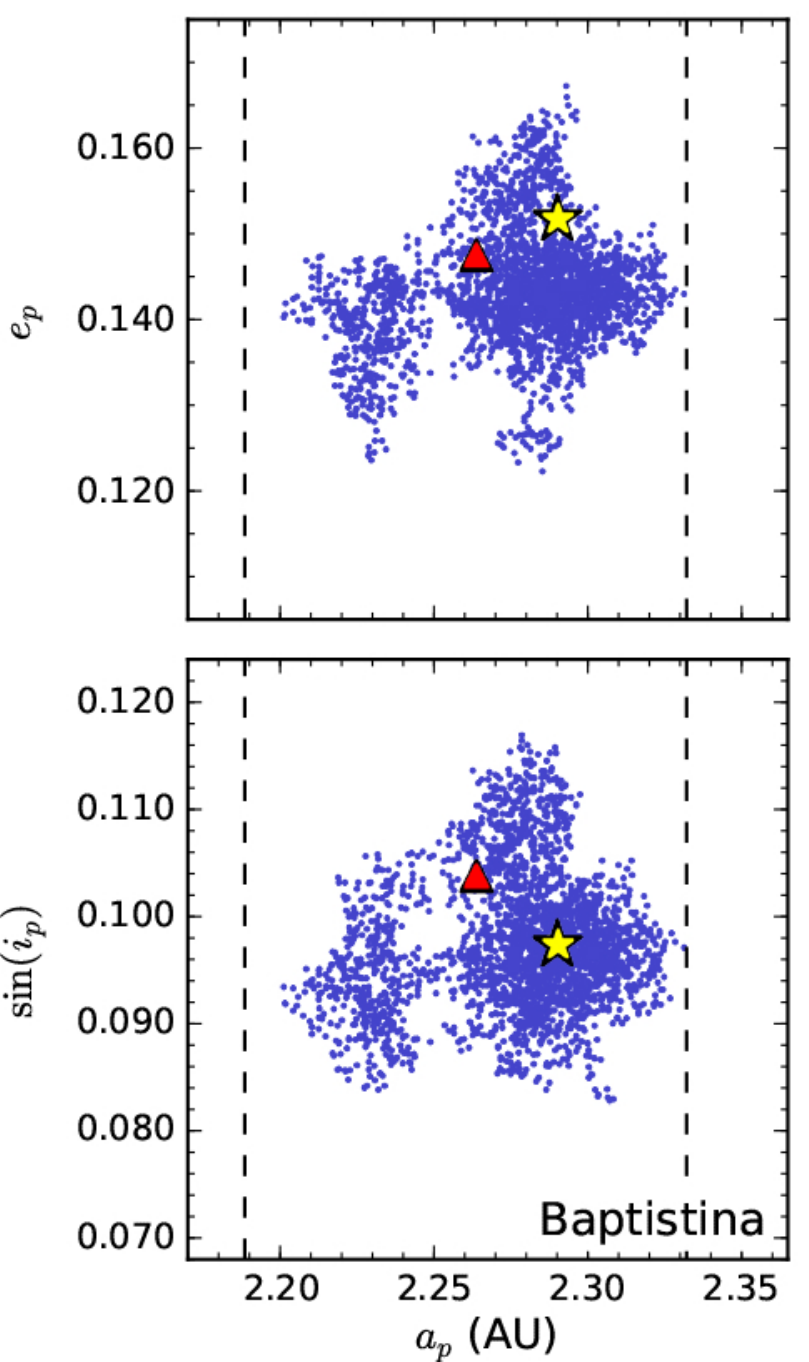}}
\caption{\small Plots of $a_p$ versus $e_p$ (top panel) and $\sin(i_p)$ (bottom panel) for Baptistina family members (small blue dots) identified by \citet{nesvorny2015_pdsastfam}.  The proper elements for (298) Baptistina are marked with red triangles, while the proper elements for 354P are marked with yellow stars.
Vertical dashed lines mark the semimajor axis positions of the 11J:3A (left) and 10J:3A (right) MMRs at 2.1885~AU and 2.3321~AU, respectively.
}
\label{figure:aei_baptistina}
\end{figure}

The Baptistina family is located in a crowded region of the inner main belt near several other families including the Flora, Vesta, Massalia, and Nysa-Polana families \citep{dykhuis2014_flora}.  Notably, the Chicxulub impactor responsible for the Cretaceous/Tertiary (K/T) mass extinction on Earth was linked to the Baptistina family by \citet{bottke2007_baptistina}, although more recent studies revising the age and composition of the family have cast doubt on this claim \citep[e.g.,][]{reddy2009_baptistina,reddy2011_baptistina,carvano2010_baptistina,masiero2012_baptistina}.  The family is roughly bound by the 11J:3A MMR on one side and the 10J:3A MMR on the other side (Figure~\ref{figure:aei_baptistina}).

The largest member of the family, (298) Baptistina, has been spectroscopically classified as a X- or Xc-type asteroid \citep{lazzaro2004_s3os2}, and \changed{has been reported to have} $p_V$$\,=\,$\changed{0.131$\,\pm\,$0.017} and $r_e$$\,=\,$\changed{10.6$\pm$0.2}~km \citep{mainzer2016_neowise}.  Other family members that have been physically characterized have been taxonomically classified as S- and X-type asteroids, and \changed{have been reported to have} $\overline{p_V}$$\,=\,$0.179$\pm$0.056 (cf.\ Table~\ref{table:family_associations}), indicating that they are not likely to have primitive compositions.

Initial analysis of 354P in 2010 indicated that the apparent cometary activity was likely to be due to a physical disruption of the asteroid by either an impact or rotational spin-up \citep{jewitt2010_p2010a2,snodgrass2010_p2010a2}, making this object a likely disrupted asteroid.  A detailed analysis of the 2010 {\it HST} images led \citet{agarwal2013_p2010a2} to conclude that the disruption of 354P was most likely due to rotational destabilization, although measurements showing the largest remaining fragment has a spin rate ($P_{rot}$$\,=\,$11.36$\pm$0.02~hr) well below the critical spin rate for rotational disruption and revised dust modeling results appear to indicate that, in fact, an impact disruption was the most likely cause of the object's observed activity in 2010 \citep{kim2017_p2010a2_1,kim2017_p2010a2_2}.  354P was the first active asteroid determined to exhibit activity that was not due to sublimation, making it the first recognized disrupted asteroid.


\subsubsection{The Behrens Family}\label{section:behrens}

\begin{figure}[htb!]
\centerline{\includegraphics[width=2.6in]{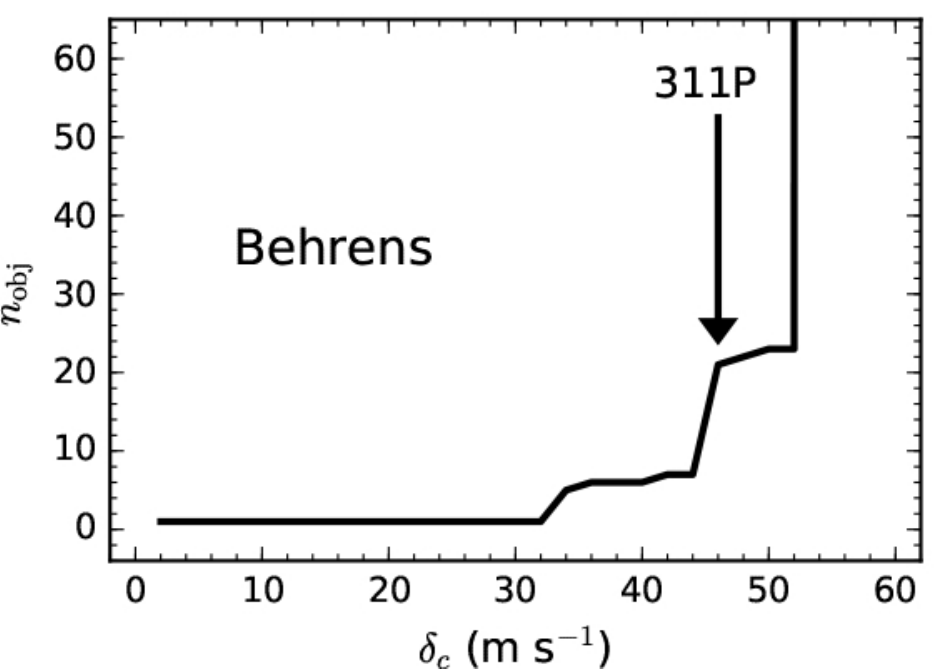}}
\caption{\small Plot of number of asteroids associated with (1651) Behrens as a function of $\delta_c$, where the point at which 311P becomes linked with the family ($\delta_c$$\,=\,$46~m~s$^{-1}$) is marked with a vertical arrow.
}
\label{figure:family_progression_behrens}
\end{figure}

We find that active asteroid 311P/PANSTARRS (formerly designated P/2013 P5) is linked to a candidate asteroid family that we designate here as the Behrens family.  311P becomes linked with the family at $\delta_c$$\,=\,$46~m~s$^{-1}$ (Figure~\ref{figure:family_progression_behrens}).  While we have not performed a detailed assessment of the likelihood that the Behrens family is real in this work, one line of evidence that the Behrens family may be real comes from the long rotational period of the asteroid (1651) Behrens, estimated to be $P_{\rm rot}$$\,\sim\,$34~hr. One possible explanation for such a long period is angular momentum ``splash'' due to a disruptive collision \citep{cellino1990_angmomentumsplash,takeda2009_collisionalspindown}. If real, the Behrens cluster is likely to be relatively young due to the small size of its largest body.  Unfortunately, this hypothesis will be difficult to confirm using the backward integration method \citep[e.g.,][]{novakovic2012_288p,novakovic2012_lorre} because the orbits of most of the family members are unstable over long timescales.

\begin{figure}[htb!]
\centerline{\includegraphics[width=2.1in]{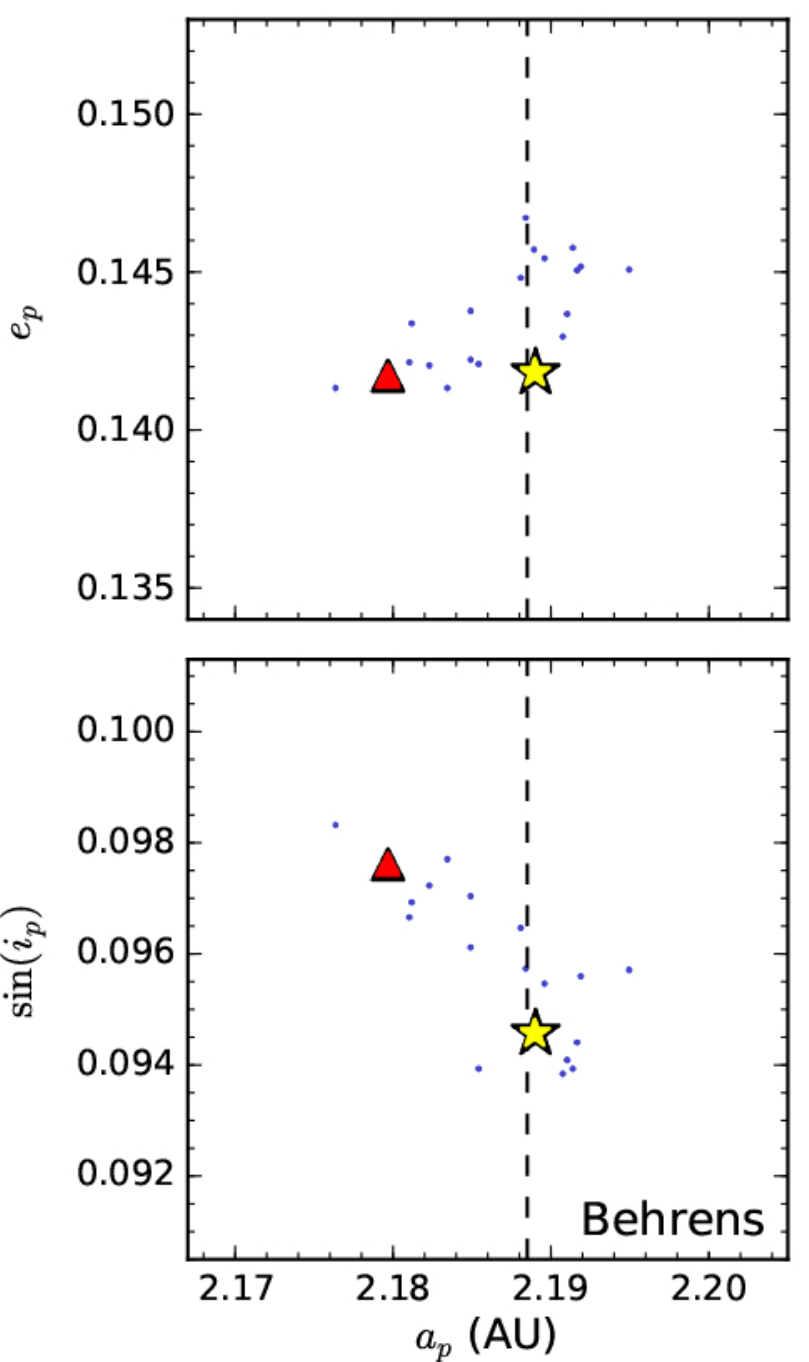}}
\caption{\small Plots of $a_p$ versus $e_p$ (top panel) and $\sin(i_p)$ (bottom panel) for candidate Behrens family members (small blue dots) identified by HCM analysis performed as part of this work using $\delta_c$$\,=\,$50~m~s$^{-1}$.  The proper elements for (1651) Behrens are marked with red triangles, while the proper elements for 311P are marked with yellow stars.  Vertical dashed lines mark the semimajor axis position of the 11J:3A MMR at 2.1885~AU.
}
\label{figure:aei_behrens}
\end{figure}

The Behrens family is intersected by the 11J:3A MMR with Jupiter.  This MMR also passes close to 311P itself (Figure~\ref{figure:aei_behrens}), suggesting that 311P may be unstable over long timescales, consistent with its relatively small $t_{ly}$ value (Table~\ref{table:aaproperties}).  As such, its membership in the Behrens family may be considered somewhat uncertain.

The largest member of this candidate family, (1651) Behrens, \changed{has been reported to have} $p_V$$\,=\,$0.318\changed{$\pm$0.052} and $r_e$$\,=\,$4.5\changed{$\pm$0.1}~km \citep{mainzer2016_neowise}, the former of which suggests that it likely does not have a primitive composition, although no formal taxonomic classification is currently available for the object.  Other family members that have been physically characterized have been taxonomically classified as Q-, S-, and V-type asteroids, and \changed{have been reported to have} $\overline{p_V}$$\,=\,$0.248$\pm$0.026 (cf.\ Table~\ref{table:family_associations}), indicating that they are not likely to have primitive compositions.

At the time of its discovery, 311P exhibited at least six dust tails believed to have been produced by multiple impulsive mass shedding events caused by rapid rotation of the nucleus near its critical limit \citep{jewitt2013_311p,jewitt2015_311p}, making the object a probable disrupted asteroid.  Attempts to confirm that the nucleus of 311P has a rapid rotation rate have thus far been unsuccessful, however, with some observations even suggesting that it could be rotating unusually slowly \citep[e.g.,][]{hainaut2014_311p}.  The nucleus has been estimated to have $r_e$$\,=\,$0.20$\pm$0.02~km \citep{jewitt2015_311p}, and has been found to have broadband colors consistent with being a S-type asteroid \citep{hainaut2014_311p}.


\subsubsection{The Gibbs Cluster}\label{section:331p}

\begin{figure}[htb!]
\centerline{\includegraphics[width=2.1in]{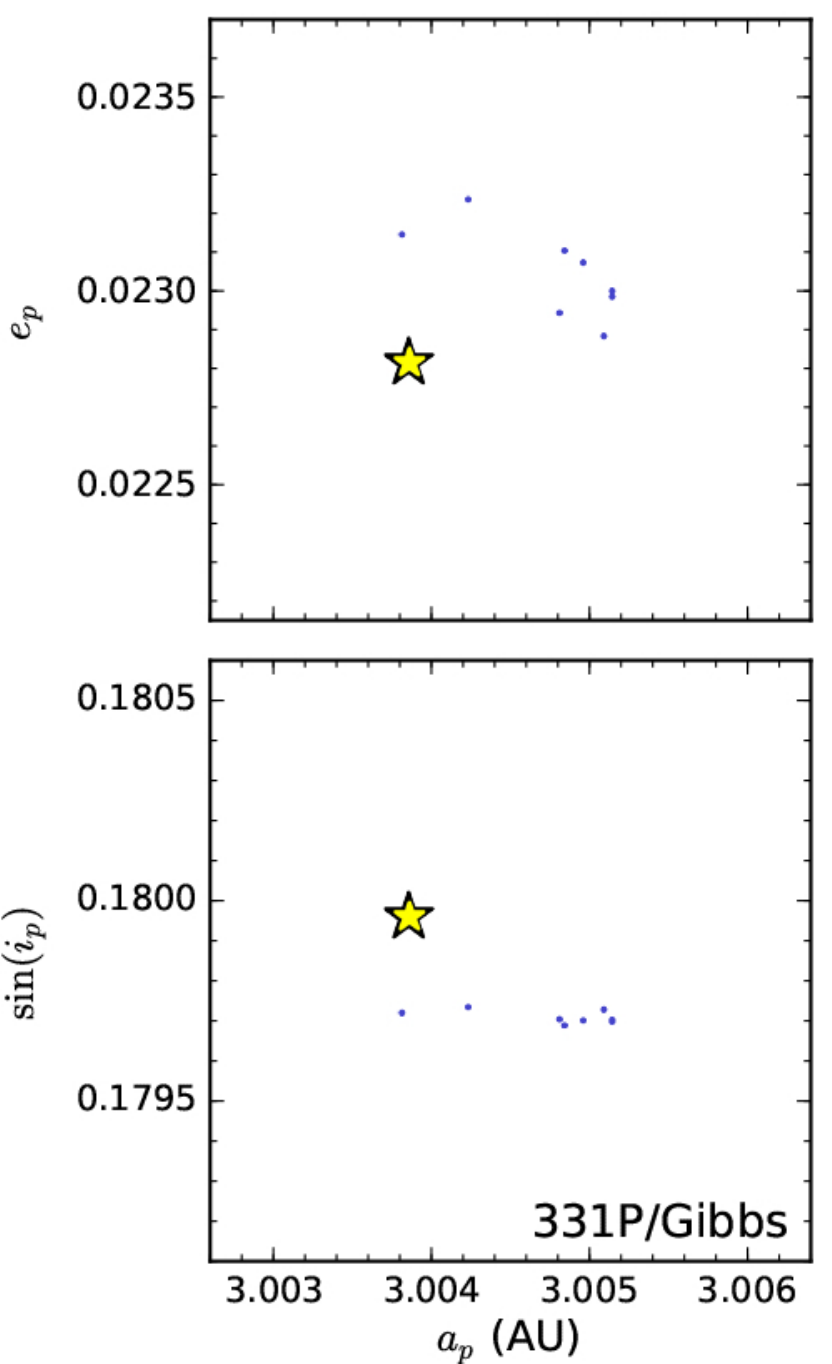}}
\caption{\small Plots of $a_p$ versus $e_p$ (top panel) and $\sin(i_p)$ (bottom panel) for 331P/Gibbs family members (small blue dots) identified by \citet{novakovic2014_331p}.  The proper elements for 331P are marked with yellow stars.
}
\label{figure:aei_gibbs}
\end{figure}

Active asteroid 331P/Gibbs (formerly designated P/2012 F5) has been previously determined to be associated with a family that was designated the Gibbs cluster and estimated to be just 1.5$\pm$0.1~Myr old \citep{novakovic2014_331p}.  
No significant two- or three-body MMRs intersect the Gibbs cluster region (Figure~\ref{figure:aei_gibbs}), making both 331P and the overall cluster relatively dynamically stable.  The parent body of the cluster has been estimated to be $\sim$10~km in diameter, where \citet{novakovic2014_331p} concluded that the estimated mass ratio between the largest fragment and the parent body indicate that the disruption that created the cluster was likely intermediate between a catastrophic disruption and a cratering event, \changed{assuming that the cluster was formed by an impact event. A more recent analysis, however, including observations to determine the rotational periods and sizes of cluster members, suggest that the cluster may instead have been formed by rotational fission \citep{pravec2017_astclusters}.}

\begin{figure}[htb!]
\centerline{\includegraphics[width=2.6in]{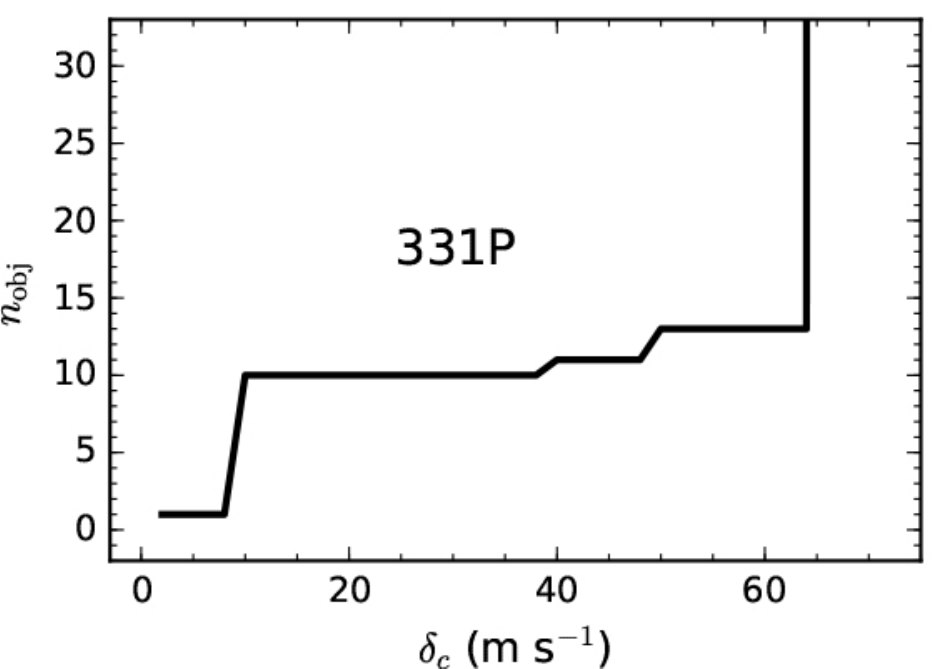}}
\caption{\small Plot of number of asteroids associated with 331P/Gibbs as a function of $\delta_c$.
}
\label{figure:family_progression_gibbs}
\end{figure}

The physical properties of this cluster were studied by \citet{novakovic2014_331p}, who noted that two members for which SDSS observations were available appear to be Q-type objects and also conducted a small-scale search for other active cluster members (none were found).
331P's nucleus has been estimated to have $r_e$$\,=\,$0.88$\pm$0.01~km \citep{drahus2015_331p} but has not yet been taxonomically classified.
Dust modeling has suggested that the long, thin dust trail observed for 331P was most likely produced by an impulsive emission event, such as an impact \citep{stevenson2012_331p,moreno2012_331p}, or possibly by mass ejection due to rotational destabilization of the nucleus given that its rotation period was found to be $P_{\rm rot}$$\,=\,$3.24$\pm$0.01~hr \citep{drahus2015_331p}, making the object a likely disrupted asteroid.


\subsubsection{The Hygiea Family}\label{section:hygiea}

\begin{figure}[htb!]
\centerline{\includegraphics[width=2.6in]{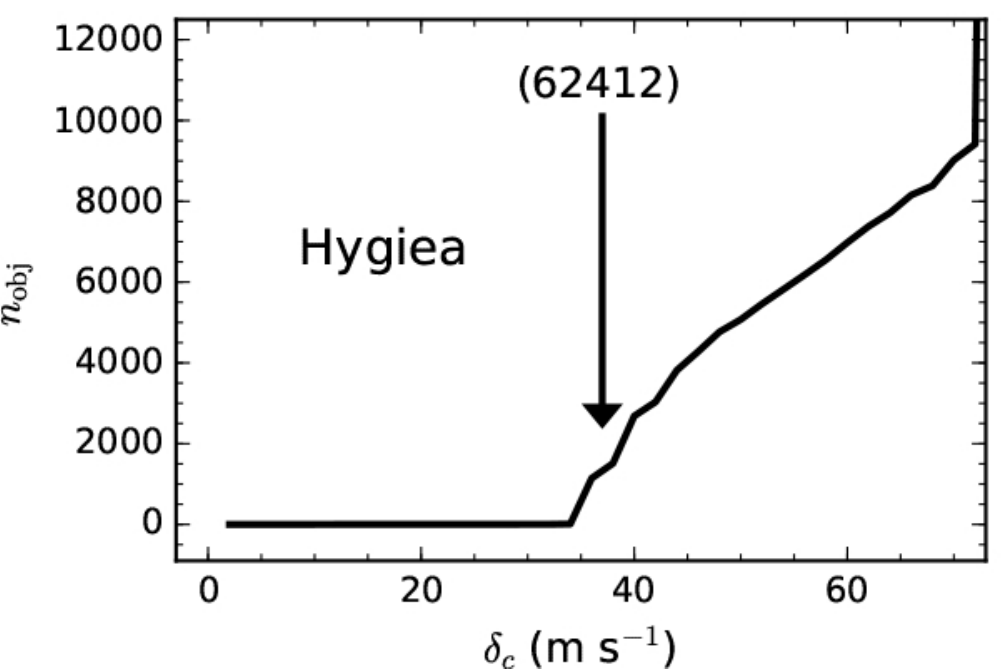}}
\caption{\small Plot of number of asteroids associated with (10) Hygiea as a function of $\delta_c$, where the point at which (62412) becomes linked with the family ($\delta_c$$\,=\,$37~m~s$^{-1}$) is marked with a vertical arrow.
}
\label{figure:family_progression_hygiea}
\end{figure}

We confirm the finding of \citet{sheppard2015_sy178} that active asteroid (62412) 2000 SY$_{178}$ is linked to the Hygiea family.
(62412) becomes linked with the Hygiea family at $\delta_c$$\,=\,$37~m~s$^{-1}$ (Figure~\ref{figure:family_progression_hygiea}), well within the optimum cut-off distance ($\delta_{c}$$\,=\,$60~m~s$^{-1}$) determined for the family by \citet{nesvorny2015_astfam_ast4}.  The Hygiea family has a size-frequency distribution consistent with being the result of the catastrophic disruption of a monolithic parent body \citep{durda2007_impactsfds,benavidez2012_sfds}, and has been determined to be 3.2$\pm$0.4~Gyr old \citep{carruba2014_hygiea}.

\begin{figure}[htb!]
\centerline{\includegraphics[width=2.1in]{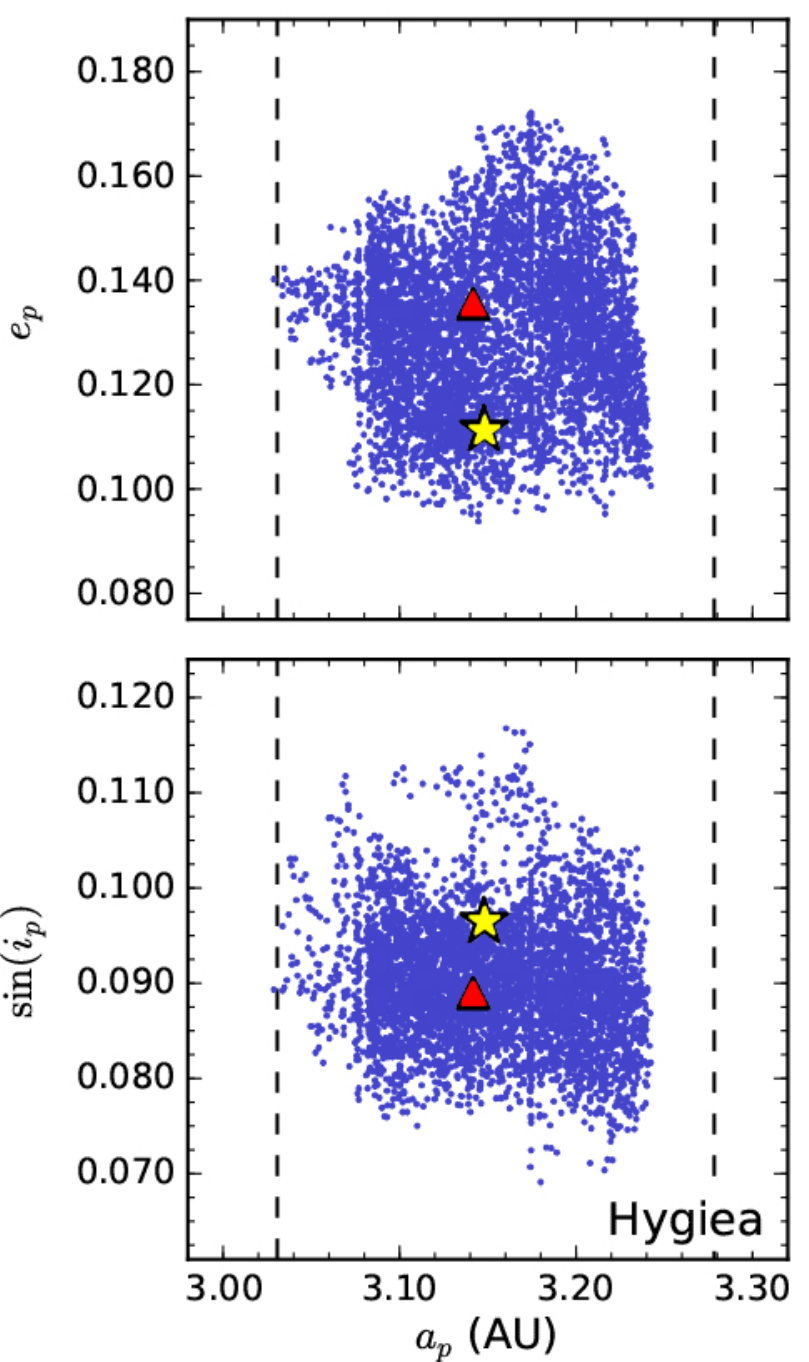}}
\caption{\small Plots of $a_p$ versus $e_p$ (top panel) and $\sin(i_p)$ (bottom panel) for Hygiea family members (small blue dots) identified by \citet{nesvorny2015_pdsastfam}.  The proper elements for (10) Hygiea are marked with red triangles, while the proper elements for (62412) are marked with yellow stars.
Vertical dashed lines mark the semimajor axis positions of the 9J:4A (left) and 2J:1A (right) MMRs at 3.0307~AU and 3.2783~AU, respectively.
}
\label{figure:aei_hygiea}
\end{figure}

The Hygiea family is roughly bound by the 9J:4A MMR at 3.0307~AU on one side and the 2J:1A MMR at 3.2783~AU on the other side (Figure~\ref{figure:aei_hygiea}).  \citet{carruba2013_hygiea} and \citet{carruba2014_hygiea} found that the region in which it is found likely contains a significant component of interlopers from the nearby Themis and Veritas families, which likely contribute low-albedo asteroids, and the Eos family, the likely origin of the few high-albedo asteroids found in the region.  It also crosses two other smaller families (associated with (5340) Burton and (15755) 1992 ET$_{5}$) in proper element space.  Besides the numerous two- and three-body resonances intersecting the region, Hygiea family members are also perturbed by numerous secular resonances, the Yarkovsky effect, and other massive asteroids, interestingly possibly including Hygiea itself \citep{carruba2014_hygiea}.

The largest member of the family, (10) Hygiea, has been spectroscopically classified as a C-type asteroid \citep{mothediniz2001_hygiea,neese2010_taxonomy}, and \changed{has been reported to have} $p_V$$\,=\,$0.072\changed{$\pm$0.002}, $r_e$$\,=\,$203.6\changed{$\pm$3.4}~km, and $\rho$$\,=\,$2.19$\pm$0.42~kg~m$^{-3}$ \citep{mainzer2016_neowise,carry2012_astdensities}.  The asteroid's spectrum includes an absorption feature centered at 3.05$\pm$0.01~$\mu$m that has been classified as ``Ceres-like'' by \citet{takir2012_3micron}.  The corresponding feature on Ceres may be due to irradiated organic material and crystalline water ice, or perhaps iron-rich clays \citep{vernazza2005_ceresvesta,rivkin2006_ceres}.  Rotationally resolved spectroscopy of Hygiea has also revealed surface heterogeneity suspected of being due to heating by significant impact events \citep{busarev2011_spectralheterogeneity}.  Other family members have been taxonomically classified as B-, C-, D-, S-, V-, and X-type asteroids (some of which may be interlopers) \citep{mothediniz2001_hygiea,carruba2013_hygiea,carruba2014_hygiea}, and \changed{have been reported to have} $\overline{p_V}$$\,=\,$0.070$\pm$0.018 (cf.\ Table~\ref{table:family_associations}), indicating that most members are likely to have primitive compositions.

Dust emission observed from asteroid (62412) 2000 SY$_{178}$ in 2014 is considered likely to have been driven by rotational disruption, given the determination of a relatively rapid rotational period for the object of $P_{\rm rot}$$\,\sim\,$3.33~hr \citep{sheppard2015_sy178}, making it a likely disrupted asteroid.  The object's nucleus has been estimated to have an effective radius of $r_e$$\,=\,$3.9$\pm$0.3~km with a minimum axis ratio of $a/b$$\,\geqslant\,$1.51, where measured colors and low albedo suggest that it is a C-type asteroid \citep{sheppard2015_sy178}.

\subsection{Active Asteroids Without Associated Families}\label{section:nofamilies}

We do not find any families associated with active asteroids 233P/La Sagra, 259P/Garradd, 348P/PAN-STARRS, (1) Ceres, (596) Scheila, and (493) Griseldis.  Of these objects, 259P, 348P, and Ceres have exhibited likely sublimation-driven activity, Scheila's activity was caused by an impact disruption, and the sources of activity exhibited by 233P, 348P, and Griseldis have yet to be determined.

259P/Garradd was first observed to be active in 2008 \citep{jewitt2009_259p}, and recently confirmed to exhibit recurrent activity \citep{hsieh2017_259p}, strongly suggesting that its activity is sublimation-driven.  The object's nucleus has $r_e$$\,=\,$0.30$\pm$0.02~km \citep[assuming $p_R$$\,=\,$0.05;][]{maclennan2012_259p}.  Dynamically, 259P has been determined to be unstable on a timescale of $\sim$20$-$30~Myr, indicating that it is unlikely to be native to its current orbit, and may have instead originated elsewhere in the main belt or possibly as a JFC \citep{jewitt2009_259p,hsieh2016_tisserand}.

\begin{figure*}[tbp]
\centerline{\includegraphics[width=5in]{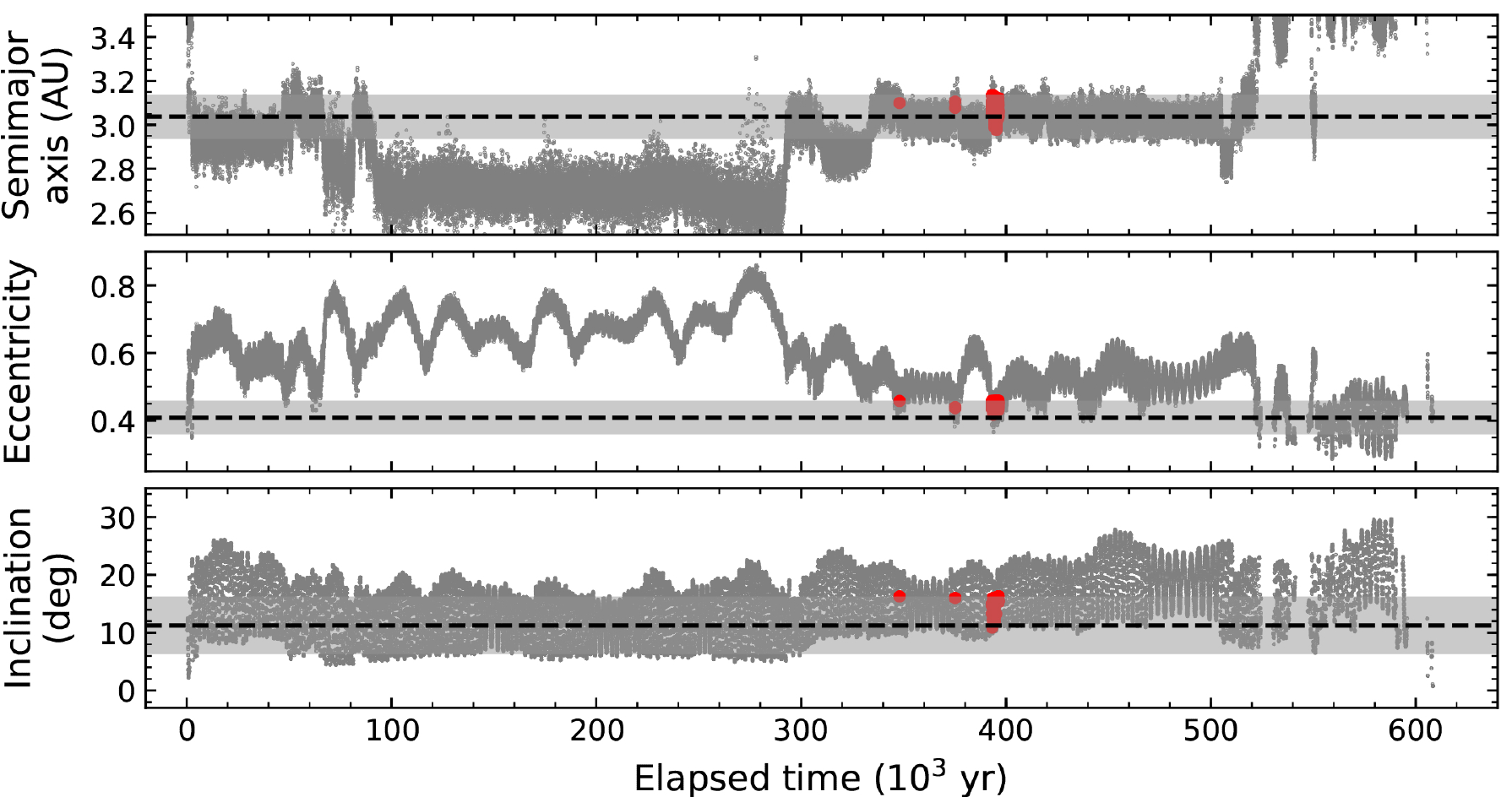}}
\caption{\small Plots of semimajor axis in AU (top panel), eccentricity (middle panel), and inclination in degrees (bottom panel) as a function of time (small grey dots) for a synthetic Jupiter-family comet from \citet{brasser2013_oortsdformation}.  Regions shaded in light grey indicate where orbital elements are similar to those of 233P, specifically where $a$$\,=\,$$a_{\rm 233P}\pm0.1$~AU, $e$$\,=\,$$e_{\rm 233P}\pm0.05$, and $i$$\,=\,$$i_{\rm 233P}\pm5^{\circ}$, where $a_{\rm 233P}$, $e_{\rm 233P}$, and $i_{\rm 233P}$ are the semimajor axis in AU, eccentricity, and inclination in degrees of 233P, respectively.  Red dots indicate where the orbital elements of the synthetic comet simultaneously meet all of these criteria for being similar to 233P's orbital elements. }
\label{figure:233p_evolution}
\end{figure*}

233P was discovered to be active by the WISE spacecraft \citep{mainzer2010_233p}.  Aside from a small number of ground-based observations confirming the presence of activity for the discovery announcement, no follow-up observations have been published to date.  As such, little is known about the object's physical properties while either active or inactive, and no assessment about the likely cause of its activity is currently available.  Its relatively high eccentricity ($e$$\,=\,$0.409), perihelion distance ($q$$\,=\,$1.795~AU) close to the aphelion of Mars ($Q_{\rm Mars}$$\,=\,$1.666~AU), and $T_J$ value ($T_J$$\,=\,$3.08) in the indistinct dynamical boundary region between asteroids and comets \citep[cf.][]{hsieh2016_tisserand} suggest, however, that it may simply be a Jupiter-family comet (JFC) that has briefly taken on main-belt-like orbital elements.  This conclusion is supported by the object's small $t_{ly}$ (Table~\ref{table:aaproperties}) as well as the fact that we find that several synthetic JFCs studied by \citet{brasser2013_oortsdformation} take on 233P-like orbital elements at some point during their evolution.  We plot the orbital evolution of an example of such an object in Figure~\ref{figure:233p_evolution}.

348P/PANSTARRS was discovered in 2017 \citep{wainscoat2017_p2017a2}.  It has a small $t_{ly}$ value (Table~\ref{table:aaproperties}), indicating that it is dynamically unstable, and its semimajor axis ($a$$\,=\,$3.166~AU) is also close to the 19J:9A MMR at 3.1623~AU.  Like 233P, it has a relatively high eccentricity ($e$$\,=\,$0.301) for a main-belt asteroid and also has $T_J$$\,=\,$3.062, placing it within the dynamical boundary region between asteroids and comets \citep[cf.][]{hsieh2016_tisserand}. We also find that two synthetic JFCs studied by \citet{brasser2013_oortsdformation} briefly take on 348P-like orbital elements during their evolution.  As such, we suspect that it may also be a JFC that has temporarily taken on main-belt-like orbital elements.

For completeness, we include dwarf planet (1) Ceres as an active asteroid given that water vapor has been detected from the body
by the {\it Herschel Space Observatory} \citep{kuppers2014_ceres}.  Of course, due to its much larger size \citep[$r_e$$\,\sim\,$470~km;][]{carry2008_ceres,park2016_ceres} relative to the other objects we are considering here, the physical regime occupied by the object is certainly very different from those occupied by other active asteroids.  No family has been identified for Ceres to date \citep[e.g.,][]{milani2014_astfamilies,rivkin2014_ceresfamily}, although \citet{carruba2016_ceresfamily} proposed that Ceres family members might simply be highly dispersed and therefore undetectable by standard family identification techniques.

Scheila was observed to be active in 2010, exhibiting an unusual three-tailed morphology \citep{jewitt2011_scheila,bodewits2011_scheila}.  Scheila's activity was most likely due to an oblique impact which generated an impact cone and down-range plume of impact ejecta \citep{ishiguro2011_scheila2}.  The asteroid has been classified as a T-type asteroid \citep{neese2010_taxonomy}, and \changed{has been reported to have} $p_V$$\,=\,$0.040\changed{$\pm$0.001} and $r_e$$\,=\,$79.9\changed{$\pm$0.6}~km \citep{mainzer2016_neowise}.  Dust emission observed from (493) Griseldis in 2015 was likewise suspected of being impact-generated, due to the short duration of the observed activity and morphology of the detected extended dust feature \citep{tholen2015_griseldis}, although a detailed analysis of its activity has yet to be published.  Griseldis has been classified as a P-type asteroid \citep{neese2010_taxonomy} and \changed{has been reported to have} \changed{$p_V$$\,=\,$0.081$\pm$0.009 and} $r_e$$\,=\,$20.8\changed{$\pm$0.1}~km \citep{mainzer2016_neowise}.  We do not find any families associated with Scheila or Griseldis.

\subsection{Other Asteroid Families and Clusters}\label{section:family_other}

There are some young asteroid families with which no known active asteroids are currently associated, but have properties suggesting that they could be found in the future to contain active asteroids.  The Veritas family
has been determined to be 8.3$\pm$0.5~Myr old \citep{nesvorny2003_dustbands} and is dominated by C-type asteroids \citep{mothediniz2005_familyspectroscopy}.
No MBCs have been associated with this family to date, although its young age and primitive composition strongly suggests that it could have the potential to harbor them \citep[cf.][]{hsieh2009_htp}.

Another interesting group of asteroids is the Lorre cluster, named for (5438) Lorre,
which was determined to be 1.9$\pm$0.3~Myr old by \citet{novakovic2012_lorre}.  Lorre has been classified as a C-type asteroid, and is the only object in the 19-member cluster to have had its spectral class determined.  The average \changed{reported} albedo of ten cluster members for which albedos have been measured is $\overline{p_V}$$\,=\,$\changed{0.044$\pm$0.013}, though, consistent with these other members also being C-type objects.  No MBCs have yet been identified among the members of the cluster, although due to the cluster's young age and spectral type of its largest member, \citet{novakovic2012_lorre} hypothesized that it could be a potential MBC reservoir.

In these cases of young primitive asteroid families for which no MBCs have yet been found, the lack of currently known MBCs in these families could be due to the fact that not all members of these families have been observed deeply enough or at the right times to reveal faint, transient cometary activity \citep[cf.][]{hsieh2009_htp}.  Impact-triggered activation of MBC activity also depends on the local collision rate and so the rate of activations may simply be lower in certain families, particularly those at higher inclinations \citep[e.g.,][]{farinella1992_astcollisionrates}.

In addition to the young asteroid families discussed in this section and earlier in this paper, a number of other young asteroid families or clusters, including the Datura, Brugmansia, Emilkowalski, Hobson, Iochroma, Irvine, Kap'bos, Lucascavin, Nicandra, Rampo, and Schulhof families, all of which have ages of $<$$\,$2~Myr, have been identified \citep{nesvorny2015_astfam_ast4,pravec2017_astclusters}.  The \changed{reported} albedos of most of the central bodies of these families for which albedos have been measured are large \citep[$p$$\,>\,$0.1;][]{mainzer2016_neowise}, however, suggesting these families do not have particularly primitive compositions, and so are unlikely to contain MBCs.  One exception is (66583) Nicandra, which \changed{has been reported to have} $p_V$$\,=\,$0.049\changed{$\pm$0.007} \citep{mainzer2016_neowise}, suggesting that the members of its associated family could be primitive and therefore potentially icy.


\section{Discussion}\label{section:discussion}

\subsection{Overall Results}\label{section:discussionoverview}

As can be seen from Table~\ref{table:family_associations}, nearly all of the known active asteroids or active asteroid fragments we considered appear to be associated with at least one asteroid family.  Of those objects found to not have family associations, 233P, 259P, and 348P have been dynamically determined to be potential interlopers at their present locations (Section~\ref{section:nofamilies}), P/2013 R3-A has a corresponding fragment (P/2013 R3-B) that is associated with a family, where both fragments are closely associated with a significant MMR (9J:4A) (Section~\ref{section:mandragora}), Ceres has exhibited volatile outgassing, but is a very large object and so occupies a very different physical regime from other much smaller suspected MBCs, and Griseldis and Scheila are also relatively large objects that are suspected of undergoing impact disruptions (Section~\ref{section:nofamilies}).

Of the 384\,337 asteroids used for the analysis performed by \citet{nesvorny2015_astfam_ast4} (including Hungaria, Hilda, and Jovian Trojan asteroids), $\sim$143\,000 were linked to 122 families, corresponding to an average family association rate of $\sim$37\%.  This number notably omits members of 19 groupings designated as candidate families (of which, for example, the cluster associated with 288P is one) by the authors due to those groupings' uncertain statistical significance at the time, and other families may have yet to be identified, \changed{and so represents a lower limit to the true combined family or candidate family association rate for inner solar system asteroids.}  Nonetheless, even including outliers in physical size like Ceres, Griseldis, and Scheila, and possible interlopers like 259P, 233P, and 348P, we find a family or candidate family association rate for active asteroids and active asteroid fragments of 16 out of 23, or $\sim$70\%, significantly higher than the currently known ``background'' family association rate.

Treating the fragments of P/2013 R3 and P/2016 J1 as non-independent objects, we find a family or candidate family association rate for MBCs of 10 out of 12.  Using 37\% as the average likelihood of a random asteroid being associated with a family, there is a 0.1\% probability of this family association rate occurring by pure chance.  
Meanwhile, for disrupted asteroids, there is a $\sim$6\% probability of seeing the observed family association rate (5 out of 7) by pure chance.
Of course, if the aforementioned physical and dynamical outliers are removed from consideration, we would then find an even higher MBC family or candidate family association rate, making the likelihood that the observed family association rate has occurred by chance even more improbable.


While in most cases, physical information is available for only a small fraction of the members of each family associated with an active asteroid, we find that all asteroid families associated with MBCs contain at least some primitive objects, i.e., objects taxonomically classified as C-complex or even D-type asteroids (Table~\ref{table:family_associations}), where the few MBCs that have been directly taxonomically classified are also all found to have C-complex spectra.  This result is likely related to the higher prevalence of primitive-type asteroids in the outer main belt \citep[cf.][]{demeo2015_astbeltcomposition_ast4} where most MBCs are found \citep[cf.][]{hsieh2015_ps1mbcs}, but may also be significant on its own, given that a MBC in the middle main belt (P/2015 X6) is also associated with a family with C-type asteroids in it (the Aeolia family).  Meanwhile, the taxonomic types of members of families associated with disrupted asteroids are more diverse, including both primitive C-complex and D-type asteroids as well as less primitive Q-, S-, and V-type asteroids (Table~\ref{table:family_associations}).

With just 23 active asteroid and active asteroid fragments considered, the statistical significance of the analysis presented here is certainly limited by our small sample size. \changed{The average family association rate for asteroids in the inner solar system is likely also dependent on the specific region being considered and possibly also the taxonomic types of the objects being considered.  For example, it might be more appropriate for us to interpret our MBC family association rate using the average family association rate for primitive-type asteroids in the outer main belt, which is likely to be different from the family association rate for the entire population of main belt, Hungaria, Hilda, and Jovian Trojan asteroids considered by \citet{nesvorny2015_astfam_ast4}.  Unfortunately, we lack the orbital element distribution of the specific catalogue of asteroids used by \citet{nesvorny2015_astfam_ast4} as well as the taxonomic classifications for the vast majority of small asteroids that are necessary to derive average family association rates taking those properties into account.  We also note that if every main-belt asteroid were subjected to the same dynamical scrutiny as each active asteroid, more might be found to be associated with their own candidate families.  Addressing these various uncertainties in the average asteroid family association rate used to evaluate the significance of our observed MBC family association rate is well beyond the scope of this work.  For reference though, we note that using 50\% as the average asteroid family association rate, there is a 1.6\% probability of our observed MBC family association rate occurring by chance, and using 90\% as the average asteroid family association rate, there is a 23\% probability of our observed MBC family association rate occurring by chance.  Meanwhile, there are 16\% and 12\% probabilities of seeing our observed disrupted asteroid family association rate given average asteroid family association likelihoods of 50\% and 90\%, respectively.}


\subsection{Implications for MBCs}\label{section:implications_mbcs}






\subsubsection{Finding and Characterizing New MBCs}\label{section:mbc_dicovery}

From the perspective of finding new MBCs, because asteroid family members are thought to have similar compositions \citep[e.g.,][]{ivezic2002_astfamilies,vernazza2006_karin}, it is logical to expect that asteroid families that contain a known, presumably ice-bearing MBC could contain other icy objects.  It is this hypothesis that led \citet{hsieh2009_htp} to search members of the Themis family (already known to contain 133P) for new MBCs, ultimately leading to the discovery of activity for asteroid (118401) LINEAR, now also designated as 176P.  
Meanwhile, dynamical studies have also shown that intra-family collision rates may be elevated over local background rates, particularly for young families \citep[e.g.,][]{farinella1992_astcollisionrates,delloro2002_newfamilycollisionrates}, suggesting that more potentially activity-triggering impacts on icy objects might occur in asteroid families, further increasing the chances of producing active MBCs.

In terms of characterizing both new and known MBCs, many MBC nuclei have been found to be quite small, with effective radii of $r_e$$\,\lesssim\,$1~km (cf.\ Table~\ref{table:aaproperties}).  Such objects are difficult to physically characterize as they are extremely faint \citep[e.g., $m_R$$\,\sim\,$24-26~mag;][]{maclennan2012_259p,hsieh2014_324p} when inactive far from the Sun, making reliable photometry, colors, or spectroscopy difficult to obtain at those times.  Meanwhile, their surface properties also cannot be measured when they are closer to the Sun and therefore brighter, as this is where they become active and thus become obscured by coma dust.  In these cases, identification of an associated asteroid family can allow for reasonable guesses of a MBC's taxonomic type and albedo by proxy using corresponding measurements of other asteroids in the same family.

The Themis family provides an illustrative example of this application of establishing links between MBCs and specific asteroid families as it contains two MBCs (133P and 176P) that have been individually characterized as B- or F-type asteroids and \changed{have been reported to have} albedos of $p_R$$\,=\,$0.05\changed{$\pm$0.02} and $p_R$$\,=\,$0.06\changed{$\pm$0.02} \citep{hsieh2009_albedos}, where the family as a whole has been found to contain mostly C-complex asteroids (which include B- and F-type asteroids) and \changed{has been reported to have} an average albedo of $\overline{p_V}$$\,=\,$0.068$\pm$0.017 (Section~\ref{section:themis_beagle}; Table~\ref{table:family_associations}).  This technique for inferring a MBC's taxonomic type is obviously subject to uncertainties due to possible differentiation of the family parent body, or if the MBC is an interloper or taxonomically dissimilar member of the background population or overlapping family.  In the absence of other direct surface property measurements of MBC nuclei, fellow family members may nonetheless be able to provide useful information about the likely properties of those MBC nuclei from their dynamical associations alone.

It is therefore interesting to note that, although only three MBC nuclei have been individually taxonomically classified (as B-, C-, or F-type asteroids) and two \changed{have been reported to have} albedos ($p_R$$\,\sim\,$0.05), all MBCs with family associations to belong to families containing primitive-type asteroids (cf.\ Section~\ref{section:discussionoverview}) that also have relatively low average \changed{reported} albedos ($\overline{p_V}$$\,\lesssim\,$0.10).
Meanwhile members of families associated with disrupted asteroids span a wider range of taxonomic types and \changed{reported} albedos (0.06$\,<\,$$\overline{p_V}$$\,<\,$0.25).  These results are consistent with MBC activity being correlated to composition (i.e., whether an object contains primitive and therefore potentially icy material) and processes that produce activity in disrupted asteroids being less sensitive to composition (although may still have some dependence, e.g., to the extent that material density can affect an object's susceptibility to rotational disruption; Section~\ref{section:implications_other}).


\subsubsection{MBC Formation}\label{section:mbc_origins}


The link between MBCs and very young asteroid families (e.g., 133P and the Beagle family; Section~\ref{section:themis_beagle}) is particularly interesting considering thermal models and impact rate calculations \citep[e.g.,][]{schorghofer2008_mbaice,prialnik2009_mbaice,hsieh2009_htp} showing that ice may become depleted from the surface of a main-belt asteroid over Gyr timescales to the point at which small ($\sim$m-scale) impactors are unable to penetrate deeply enough to trigger sublimation-driven activity \citep[cf.][]{hsieh2004_133p,capria2012_mbcactivity,haghighipour2016_mbcimpacts}.  However, if most MBC nuclei were produced in more recent fragmentation events (e.g., $\lesssim\,$10~Myr), they may have much younger effective ages than their dynamical stability timescales would otherwise suggest, and could possess more ice at shallower depths than expected.

\begin{figure}[tbp]
\includegraphics[width=2.5in]{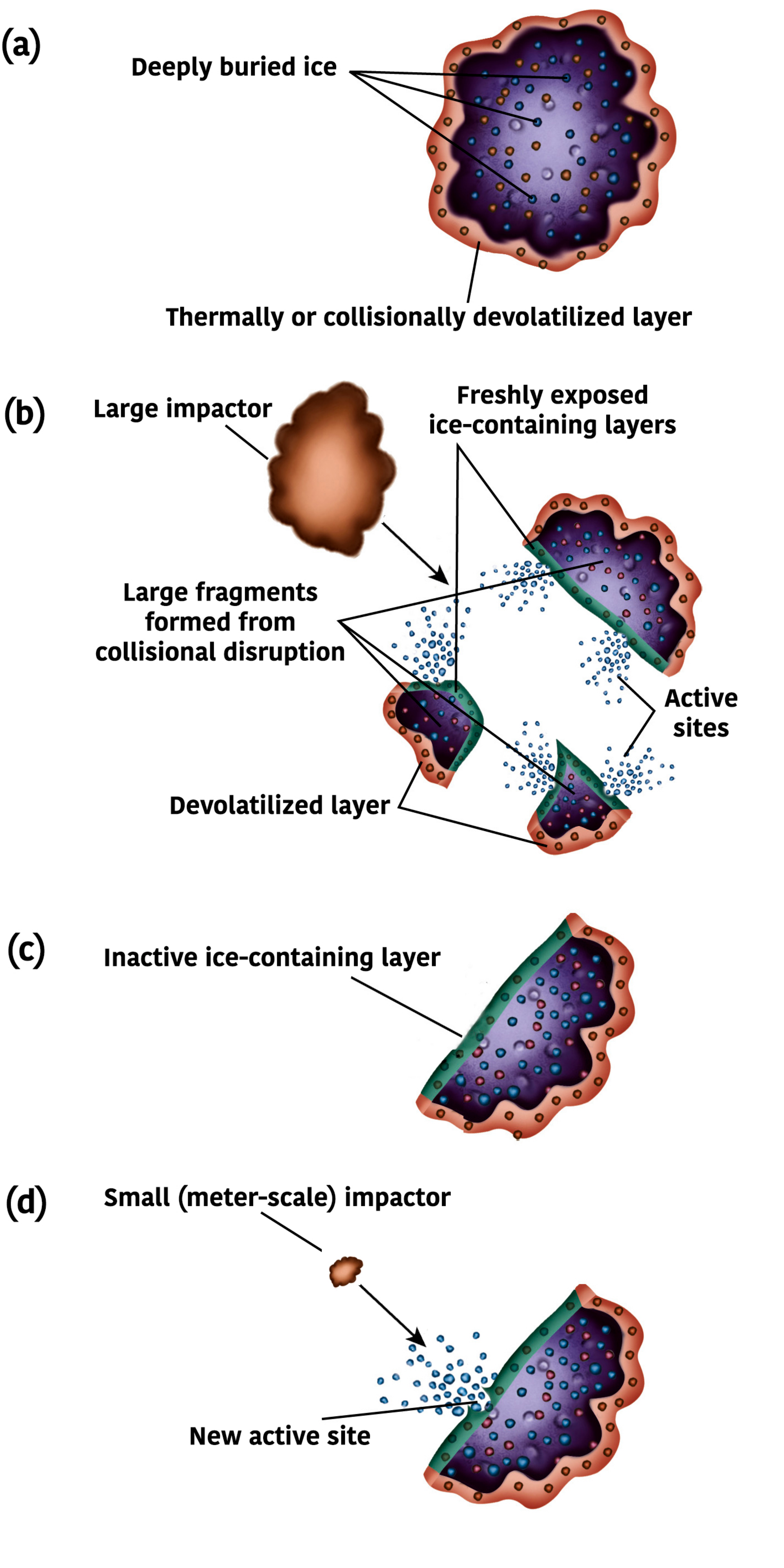}
\caption{Illustration of processes that could produce MBCs with shallow subsurface ice from the catastrophic disruption of parent bodies with initally more deeply buried ice: (a) a long-lived icy main belt object has preserved ice in its interior but has had its outer layer largely devolatilized from solar heating and impact gardening over $\sim$Gyr timescales; (b) a large impactor \changed{(or rotational instability)} catastrophically disrupts this body, exposing its icy interior and leading to sublimation-driven outbursts from ice directly exposed by the disruption; (c) mantling on new family members quenches activity triggered directly by the disruption of the family's parent body, although ice still remains relatively close to the surface; and (d) a small (and therefore abundant and relatively frequently encountered) impactor, which would otherwise be unable to penetrate the inert surface layer of an older ice-bearing asteroid, causes a small-scale disruption of the young family member's relatively fresh surface, excavating shallow subsurface ice and producing a localized active site from which sublimation-driven dust emission can occur.
}
\label{figure:family-formation-mbcs}
\end{figure}

In Figure~\ref{figure:family-formation-mbcs}, we illustrate a sequence of physical processes \citep[some of which have been previously noted in the context of MBCs; e.g.,][]{nesvorny2008_beagle,hsieh2009_htp,capria2012_mbcactivity,haghighipour2016_mbcimpacts} that could lead to active MBCs in young asteroid families. We begin with the premise that large icy asteroids can preserve ice over Gyr timescales in the main asteroid belt \citep[cf.][]{schorghofer2008_mbaice,prialnik2009_mbaice}, except that, by now, that ice has likely receded too deep below the surface to plausibly produce sublimation-driven activity (Figure~\ref{figure:family-formation-mbcs}a).
However, if one of these objects is catastrophically disrupted \changed{\citep[either by an impact event, as is commonly assumed, or a rotational fission event, as is suspected for some families by][]{pravec2017_astclusters}} (Figure~\ref{figure:family-formation-mbcs}b), after the sublimation and depletion of ice directly exposed by the initial disruption, remaining subsurface ice on the resulting fragments (i.e., family members) might then be found at much shallower depths than before (Figure~\ref{figure:family-formation-mbcs}c).  At these depths, that ice would then be more easily excavated by relatively small (and therefore relatively abundant) impactors.  For asteroids formed in recent ($\lesssim$10~Myr) fragmentation events where ice has not yet had time to recede again to significant depths, such small-scale disruptions should be able to trigger the sublimation-driven activity observed today on MBCs.


Interestingly, with some exceptions, most MBC nuclei are small (km-scale or smaller; Table~\ref{table:aaproperties}). Smaller objects are collisionally disrupted on statistically shorter timescales than larger objects \citep{cheng2004_collisionalevolution,bottke2005_collisionalevolution}, meaning that currently existing smaller objects are more likely to have been recently formed than larger objects.  Thus, even those MBCs for which young families have not yet been associated may have also formed in recent disruptions of larger parent bodies.  The current lack of identified young family associations for these MBCs could simply be due to other family members being too faint to have been discovered yet by current asteroid surveys.  As surveys improve and find ever fainter asteroids, more young families will likely be discovered and some of these will likely be associated with MBCs that currently lack identified young family associations.

We emphasize, however, that we do not expect young family associations to eventually be found for all MBCs.  Some MBCs may have formed in recent disruptions from which they are the only remaining fragments of appreciable size, or where family members have rapidly dispersed and blended beyond recognition into the background asteroid population due to chaotic dynamical conditions.  Some MBCs themselves may have been destabilized by chaotic dynamical conditions near their points of origins and are now interlopers at their present-day locations, far from their original families.  Finally, for certain orbital obliquities and latitudes of subsurface ice reservoirs, thermal modeling indicates that shallow subsurface ice could remain preserved on an outer main belt asteroid over even Gyr timescales \citep{schorghofer2008_mbaice,schorghofer2016_asteroidice}.  Therefore, in these cases, a recent catastrophic disruption would not be required at all for ice to be accessible to excavation by small impactors.

As ongoing surveys continue to discover more asteroids, and particularly as future surveys discover smaller and fainter asteroids than are detectable now, continued searches for tightly clustered young families should be performed \citep[e.g.,][]{milani2014_astfamilies}, \changed{perhaps using osculating or mean elements rather than proper elements for detecting very young clusters \citep[e.g.,][]{nesvorny2006_datura,nesvorny2006_youngfamilies,pravec2009_asteroidpairs,pravec2017_astclusters,rosaev2017_hobson}}.  Furthermore, as new young families are identified --- especially ones associated with known MBCs, believed to contain primitive asteroids, or found in the outer main belt (i.e., $a$ between the 5J:2A MMR at 2.8252~AU and the 2J:1A MMR at 3.2783~AU) --- targeted observations or at least targeted close examination of survey data of members of these young families should be conducted to search for new MBCs.

This schematic model also suggests that thermal modeling work on volatile preservation in asteroids should take into account the fact that many icy asteroids found in the main belt today could actually have originated from the fragmentation of larger icy parent bodies some time in the relatively recent past.  In these cases, the timescale over which volatile depletion from solar heating is expected to take place is not the age of the solar system, but rather the age of an object's associated asteroid family.  As such, we suggest that thermal models computing ice retreat depths over Gyr timescales \citep[e.g.,][]{schorghofer2008_mbaice,prialnik2009_mbaice} may overestimate depths to ice at the present day, and that ice in primitive asteroids might be found at much shallower depths (and therefore be more accessible to activity-triggering impacts) than these models might indicate.

\subsection{Implications for Disrupted Asteroids}\label{section:implications_other}

Given that impact events depend more on an object's environment rather than its composition, we might not expect that the presence of an impact-disrupted active asteroid in a family would necessarily indicate that the family could contain more.  However, impact disruptions could be more common in families in general.  Intra-family collisions (which have lower velocities and so are more likely to cause non-catastrophic disruptions) may be more likely in asteroid families relative to the local background, particularly for families with low inclinations, since their members share similar orbits \citep[cf.][]{farinella1992_astcollisionrates}.  In young families whose members still have very similar orbits, non-catastrophic impact disruptions could be even more frequent.

For rotationally disrupted asteroids, the relationship between incidence rate and membership in families is more uncertain.  Given the dependence of rotational destabilization on the composition and internal structure (particularly as they relate to density) \citep[e.g.,][]{hirabayashi2014_p2013r3,hirabayashi2014_biaxialellipsoids}, a positive correlation between membership in a particular family and the prevalence of rotational disruptions could arise if family members share similar compositions or internal structures that make them more susceptible to rotational destabilization than other non-family asteroids, particularly if that family has already been found to contain at least one other rotationally disrupted asteroid.  What is unclear is whether the variation in composition or internal structure between members of a particular family and non-family asteroids, particularly if they are of similar taxonomic type, is large enough to produce significant differences in disruption rates.  It is also unclear whether the degree of commonality in internal compositions or structures among family members plays a more significant role than other factors that also affect the likelihood of rotational disruption.  For example, the Yarkovsky-O'Keefe-Radzievskii-Paddack (YORP) effect, which may be responsible for spinning up asteroids beyond their critical limits \citep{jewitt2014_p2013r3}, is known to depend on the thermal properties of an asteroid, particularly thermal inertia, which could be similar for asteroids belonging to the same family, but it also strongly depends on asteroid size and shape, which are not particularly related to family membership \citep[cf.][]{rubincam2000_yorp,scheeres2007_yorp,golubov2016_yarkovsky}.

We conclude for now that the relationships between family membership and the probability of an asteroid experiencing an impact or rotational disruption are probably weaker than the relationship between family membership and the rate of occurrence of MBCs, especially for young asteroid families largely consisting of primitive asteroids.  Nonetheless, those relationships may not be entirely negligible and are likely worth further investigation in the future using both theoretical approaches (e.g., calculation or modeling of expected impact and rotational disruption rates within and outside families in different regions of the asteroid belt) and observational approaches (e.g., continuing to note whether newly discovered disrupted asteroids are associated with young families and tallying relative occurrence rates of disruptive events within and outside those families).


\subsection{Other Considerations and Challenges}\label{section:futurework}

Two active asteroids (311P and 354P) were initially identified as members of the Flora family, where both objects are considered disrupted asteroids \citep{jewitt2010_p2010a2,jewitt2013_311p,snodgrass2010_p2010a2}, although both of these objects have since been determined to belong to other nearby families (Baptistina and Behrens, respectively; Sections~\ref{section:baptistina} and \ref{section:behrens}).  This confusion points to an issue that will assuredly continue to arise in the future as more active asteroids are found in dense regions of the main asteroid belt in orbital element space.  In the case of the Flora family, its large size and diffuse dynamical structure necessitates care in determining its true extent and membership, given the presence of several nearby neighboring families in orbital element space (i.e., the Vesta, Baptistina, Massalia, and Nysa-Polana families).  Overlapping dynamical families, which can result when families arise from parent bodies that happened to share similar orbital elements, particularly those positioned near chaos-inducing MMRs, can result in complications in attempting to infer the composition of other family members from sublimation-driven activity observed from apparently linked active asteroids.

In studies of particularly dense regions of the asteroid belt, color and albedo information can be used to help to separate true members of a family from background objects \citep[e.g.,][]{reddy2011_baptistina,dykhuis2014_flora}, though the possibility of taxonomic diversity within a family means that this technique needs to be used with care \citep[e.g.,][]{oszkiewicz2015_flora}.  
This approach will also be unhelpful for distinguishing family members in regions where background objects and family members are compositionally similar, or for which compositional information is largely unavailable \citep[cf.][]{novakovic2012_288p}.  In these cases, more careful dynamical analyses, such as selective backward integration \citep[SBIM;][]{novakovic2012_288p} to identify clusterings of secular angles in the past for subsets of family member candidates, will be useful for identifying true family members.

Lastly, as seen throughout Section~\ref{section:results}, many families that are found to contain potentially ice-bearing objects are located near or are intersected by MMRs.  In cases where a family either contains a known MBC or at least primitive-type asteroids and is significantly affected by one or more MMRs (e.g., the Adeona family and the 8J:3A MMR, and the Themis family and the 2J:1A MMR; Sections~\ref{section:adeona} and \ref{section:themis_beagle}), the nearby or intersecting MMRs could provide a means for dispersing ice-bearing family asteroids throughout the asteroid belt and beyond.  Such a process was proposed for the origin of 238P, which \citet{haghighipour2009_mbcorigins} suggested was originally a member of the Themis family that had its eccentricity driven up by that family's close proximity to the 2J:1A MMR.  Similarly, \citet{jewitt2009_259p} suggested, based on its asteroid-like $T_J$ value, that 259P could have originated elsewhere in the asteroid belt.  In this light, future dynamical studies to ascertain whether MBCs without currently recognized links to young families may have originated elsewhere in the asteroid belt will be useful for ensuring proper interpretation of the spatial distribution of the population of icy bodies in the asteroid belt and ascertaining the degree to which they can be used to trace the distribution of ice in the primordial solar system \citep[cf.][]{hsieh2014_mbcsiausproc}.

\section{Summary}\label{section:summary}
In this work, we present the following key findings:
\begin{enumerate}
\item{We report newly identified family associations between active asteroids 238P/Read and the Gorchakov family, 311P/PANSTARRS and the Behrens family, 324P/La Sagra and the Alauda family, 354P/LINEAR and the Baptistina family, P/2013 R3-B (Catalina-PANSTARRS) and the Mandragora family, P/2015 X6 (PANSTARRS) and the Aeolia family, P/2016 G1 (PANSTARRS) and the Adeona family, and P/2016 J1-A/B (PANSTARRS) and the Theobalda family.  The Gorchakov and Behrens families are candidate families identified by this work and will require further investigation to confirm that they are real families.
}
\item{We find that 10 out of 12 MBCs and 5 out of 7 disrupted asteroids are linked with known or candidate families, rates that have $\sim$0.1\% and $\sim$6\% probabilities, respectively, of occurring by chance, given an overall average family association rate of 37\% for asteroids in the inner solar system.
}
\item{All MBCs with family associations are found to belong to families that contain and are sometimes dominated by primitive-type asteroids, and have relatively low average \changed{reported} albedos ($\overline{p_V}$$\,\lesssim\,$0.10).  Meanwhile, disrupted asteroids are found to belong to families that span wider ranges of taxonomic types (including Q-, S-, and V-types) and a wider range of average \changed{reported} albedos (0.06$\,<\,$$\overline{p_V}$$\,<\,$0.25).  These findings are consistent with hypotheses that MBC activity is closely tied to an object's composition (namely whether it is likely to contain preserved ice) while processes that produce disrupted asteroid activity are less sensitive to composition.
}
\item{We describe a sequence of processes that could produce MBCs that involves the preservation of ice over Gyr timescales within large parent bodies that are subsequently catastrophically disrupted in family forming collisions in the recent past, where the resulting young family members may possess subsurface ice at relatively shallow depths and are thus more susceptible to activation than older icy asteroids.
We suggest that as ongoing surveys discover more asteroids and future surveys discover smaller and fainter asteroids, associations with young families may eventually be found for some families that currently lack such associations, though we do not expect all MBCs to eventually be found to have such associations.
}
\item{Though we also find a suggestively high rate of disrupted asteroids with associated asteroid families, the connection between asteroid families and disrupted asteroids is less clear than for MBCs.  Further theoretical and observational work is needed to clarify the significance of the rate of family associations that we find for disrupted asteroids.
}
\end{enumerate}


\acknowledgments
HHH acknowledges support from the NASA Solar System Observations program (Grant NNX16AD68G).  
BN acknowledges support by the Ministry of Education, Science and Technological Development of the Republic of Serbia, Project 176011.
RB is grateful for financial support from JSPS KAKENHI (JP16K17662).
We also thank an anonymous reviewer for useful suggestions that helped to improve this paper.

\bibliographystyle{aasjournal}
\bibliography{hhsieh_refs}   

\end{document}